\documentclass[aps,prc,superscriptaddress,twoside,twocolumn,nofootinbib,%
showpacs,floatfix]{revtex4}

\usepackage{amsmath,amssymb}
\usepackage{graphicx,bm}

\newcommand{\hc}{\text{H.\,c.}}

\begin{document}

\title{\boldmath Photoproduction of $\Xi$ off nucleons}

\author{K. Nakayama}%
\email{nakayama@uga.edu}

\affiliation{Department of Physics and Astronomy, University of Georgia,
Athens, GA 30602, USA}
\affiliation{Institut f\"ur Kernphysik, Forschungszentrum
J\"ulich, D-52425 J\"ulich, Germany}

\author{Yongseok Oh}%
\email{yoh@physast.uga.edu}

\affiliation{Department of Physics and Astronomy, University of Georgia,
Athens, GA 30602, USA}

\author{H. Haberzettl}%
\email{helmut@gwu.edu}

\affiliation{Center for Nuclear Studies, Department of Physics, The George
Washington University, Washington, DC 20052, USA}

\date{}

\begin{abstract}
The photoproduction reaction $\gamma N \to K K \Xi$ is investigated based
on a relativistic meson-exchange model of hadronic interactions.
The production amplitude is calculated in the tree-level approximation from
relevant effective Lagrangians, whose (coupling constant) parameters are
mostly fixed from the empirical data and/or quark models together with
SU(3) symmetry considerations.
Gauge invariance of the resulting amplitude is maintained by introducing the
contact currents by extending the gauge-invariant approach of Haberzettl for
one-meson photoproduction to two-meson photoproduction.
The role of the intermediate low-lying hyperons and of the intermediate
higher-mass hyperon resonances are analyzed in detail.
In particular, the basic features of the production of $\Xi^-(1318)$ in
$\gamma p \to K^+ K^+ \Xi^-$ and their possible manifestations
in the forthcoming experimental data are discussed.
\end{abstract}


\maketitle

\section{Introduction}

In principle, flavor SU(3) symmetry allows for the existence of as many
$\Xi$ resonances as the number of $N^*$ and $\Delta^*$ resonances
combined~\cite{SGM74}.
Despite this fact, not much is known about these resonances.
Indeed, only a dozen or so $\Xi$ have been identified so far;
among them, only two, $\Xi(1318)$ and $\Xi(1530)$, have four-star
status \cite{PDG}.
One of the reasons for this situation is that $\Xi$ hyperons, being
particles with strangeness $S=-2$, are difficult to produce having relatively
small production rates;
they can only be produced via indirect processes from the nucleon.
The production of $\Xi$ baryons were, so far, restricted mainly through the
$K^-p$ reactions~\cite{JADF83} or the $\Sigma$-hyperon induced
reactions~\cite{WA89}.
However, since the late 1980's, no significant progress has been made in
cascade spectroscopy due to the closing of the then existing kaon factories.
Recently, the CLAS Collaboration at the Thomas Jefferson National Accelerator
Facility (JLab) has initiated a cascade-physics program~\cite{Price1,Price2};
the collaboration, in particular, has established the feasibility to do
cascade baryon spectroscopy via photoproduction reactions like
$\gamma p \to K^+ K^+ \Xi^-$ and
$\gamma p \to K^+ K^+ \pi^- \Xi^0$~\cite{Price2,Price3}.%
\footnote{An earlier experiment on inclusive $\Xi$ photoproduction is
reported in Ref.~\cite{TPSC87}.}
A dedicated experiment for these reactions is currently underway;
preliminary total cross section data for the first reaction have already
been reported~\cite{Guo}.
Also, cascade physics has recently received a special interest in
connection with the search for the exotic pentaquark states.
In fact, the NA49 Collaboration \cite{NA49} has reported seeing a signal for
the pentaquark cascade $\Xi^{--}_5$.
However, to date, other experiments with much higher statistics have obtained
negative results~\cite{NegCas}.

\begin{figure}[t!]\centering
\includegraphics[width=0.4\textwidth,angle=0,clip]{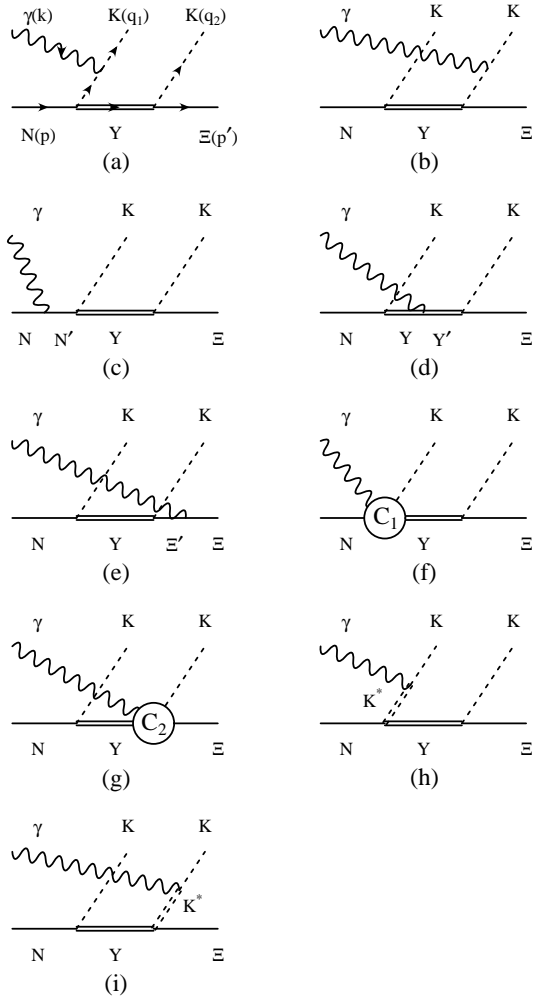}
\caption{\label{fig:diagram}
Diagrams contributing to $\gamma N \to K K \Xi$.
The intermediate baryon states are denoted as $N'$ for the nucleon and
$\Delta$, $Y,Y'$ for the hyperon $\Lambda$ and $\Sigma$ resonances,
and $\Xi'$ for $\Xi(1318)$ and $\Xi(1530)$.
The intermediate mesons in the $t$-channel are $K$ [(a) and (b)] and
$K^*$ [(h) and (i)].
The diagrams (f) and (g) are the generalized contact currents required
to maintain gauge invariance of the total amplitude.
The off-shell interaction currents corresponding to the diagram parts
$\text{C}_1$ and $\text{C}_2$, respectively, are given in
Eqs.~(\ref{eq:cnt1}) and (\ref{eq:cnt2}).
The external legs are labeled by the four-momenta of the respective particles
in (a).
Diagrams corresponding to (a)--(i) with $K(q_1) \leftrightarrow K(q_2)$
are also taken into account in the present calculation.}
\end{figure}
%

To our knowledge, the only theoretical investigation of cascade
photoproduction off nucleons is that of Ref.~\cite{Liu}, which was
devoted to the production of the pentaquark $\Xi_5^{}$ hyperon.
It is extremely timely, therefore, to study this reaction theoretically
in the energy range covered at JLab.
In the present work we investigate the $\gamma N \to K K \Xi$ reaction for
incident photon energies up to about 5 GeV.
Our approach is based on a relativistic meson-exchange model of hadronic
interactions.
The reaction amplitude is calculated in the tree-level approximation
considering the production mechanisms displayed in
Fig.~\ref{fig:diagram}, which involve excitation of baryon resonances in
the intermediate states.
In fact, the Particle Data Group (PDG)~\cite{PDG} quotes a number of four-
and three-star $\Lambda$ and $\Sigma$ hyperons which may contribute to
the production of $\Xi$ in the present reaction.
The $t$-channel meson-exchange processes with subsequent decay of the
emitted meson into two kaons (meson-production processes)
are suppressed in the present reaction, since the produced meson should be
exotic having strangeness $S=+2$.
In addition, the $t$-channel meson-exchange processes for
$\bar{K} N \to K \Xi$ are also suppressed since these can occur only via an
exchange of an exotic meson with $S=+2$.
This means that, to lowest order, the production of a cascade hyperon
involves necessarily an excitation of hyperons as shown in
Fig.~\ref{fig:diagram}.
This is quite different from the case of the $\gamma N \to K \bar{K} N$
reaction, which has large contributions from vector-meson production
processes~\cite{ONL04}.
Therefore, cascade photoproduction off nucleons should offer also a
possibility to extract information on the hyperons with $S=-1$.

In this work, we investigate the photoproduction reactions of the
ground-state cascade $\Xi(1318)$.
We will discuss four different isospin channels, namely,
\begin{equation}
\begin{aligned}
\gamma p &\to K^+ K^+ \Xi^-, & \gamma p &\to K^+ K^0 \Xi^0,
 \\
\gamma n &\to K^+ K^0 \Xi^-, & \gamma n &\to K^0 K^0 \Xi^0.
\end{aligned}
\end{equation}

At this point, we note that many issues, such as the $K\Xi$ final-state
interaction (FSI), the roles of many of the high-mass nucleon and hyperon
resonances, and high-mass vector and axial-vector mesons cannot be
addressed at this stage of the investigation due to a complete lack of
independent information as to the dynamics of how these hadrons enter the
present reaction.%
\footnote{The scalar meson exchange, however, is not present in this reaction
since the scalar meson cannot couple to the photon and the pseudoscalar meson
because of angular-momentum and parity conservation.}
To avoid any speculation on our part, we leave these issues until such time
when a better understanding of the underlying reaction dynamics is available.
Hence, it is the purpose of the present work to investigate the $\Xi$
photoproduction mechanism using only the currently available information.

In the next Section, we develop our model for $\Xi$ photoproduction,
which is shown in Fig.~\ref{fig:diagram}.
We present our results for cross sections and some spin asymmetries in
Sec.~III.
The different role of the intermediate hyperons will be discussed in
detail before we summarize in Sec.~IV.
The effective Lagrangians used in the present work and the method to
maintain the gauge-invariance condition with form factors are given
in Appendix.

\section{Model for $\bm{\gamma N \to KK\Xi}$}

In the present work, the reaction $\gamma N \to KK\Xi$ is described by the
sum of the amplitudes shown in Fig.~\ref{fig:diagram}.
The three- and four-star hyperons which may contribute to the present
reaction are summarized in Table.~\ref{tbl:hyperons}.
Among them, only for the low-mass resonances, i.e.,
$\Lambda(1116)$, $\Lambda(1405)$, $\Lambda(1520)$, $\Sigma(1190)$, and
$\Sigma(1385)$, we have sufficient information to determine the relevant
hadronic and electromagnetic coupling constants.
In fact, they can be estimated from the experimental data~\cite{PDG} and/or
from quark models and SU(3) symmetry considerations.
Tables~\ref{tbl:cc} and \ref{tbl:trcc} summarize the hyperon resonance
parameters and the corresponding estimated coupling constants.

\begin{table*}[t]
\centering
\caption{$\Lambda$ and $\Sigma$ hyperons listed by the Particle Data
Group~\cite{PDG} as three-star or four-star states.
The decay widths and branching ratios of high-mass resonances
$m_Y^{} > 1.6$ GeV are in a broad range.
The coupling constants are determined from their central values.}
\begin{tabular}{c@{\extracolsep{1em}}cccc|ccccc} \hline\hline
\multicolumn{5}{c|}{$\Lambda$ states} & \multicolumn{5}{c}{$\Sigma$ states} \\
\hline State & $J^P$ & $\Gamma$ (MeV) & Rating & $|g_{N\Lambda K}|$
& State & $J^P$ & $\Gamma$ (MeV) & Rating  & $|g_{N\Sigma K}|$ \\ \hline
$\Lambda(1116)$ & $1/2^+$ &  & **** &     &
  $\Sigma(1193)$ & $1/2^+$ &              & **** &     \\
$\Lambda(1405)$ & $1/2^-$ & $\approx 50$ & **** &       &
  $\Sigma(1385)$ & $3/2^+$ & $\approx 37$ & **** &      \\
$\Lambda(1520)$ & $3/2^-$ & $\approx 16$ & **** &       &
                 &         &              &     &       \\
\hline
$\Lambda(1600)$ & $1/2^+$ & $\approx 150$ & *** &  4.2 &
  $\Sigma(1660)$ & $1/2^+$ & $\approx 100$ & *** & 2.5  \\
$\Lambda(1670)$ & $1/2^-$ & $\approx 35$ & **** &  0.3 &
  $\Sigma(1670)$ & $3/2^-$ & $\approx 60$ & **** & 2.8 \\
$\Lambda(1690)$ & $3/2^-$ & $\approx 60$ & **** &  4.0 &
  $\Sigma(1750)$ & $1/2^-$ & $\approx 90$ & *** & 0.5 \\
$\Lambda(1800)$ & $1/2^-$ & $\approx 300$ & *** &  1.0 &
  $\Sigma(1775)$ & $5/2^-$ & $\approx 120$ & **** & \\
$\Lambda(1810)$ & $1/2^+$ & $\approx 150$ & *** &  2.8 &
  $\Sigma(1915)$ & $5/2^+$ & $\approx 120$ & **** & \\
$\Lambda(1820)$ & $5/2^+$ & $\approx 80$ & **** &      &
  $\Sigma(1940)$ & $3/2^-$ & $\approx 220$ & *** & $< 2.8$  \\
$\Lambda(1830)$ & $5/2^-$ & $\approx 95$ & **** &      &
  $\Sigma(2030)$ & $7/2^+$ & $\approx 180$ & **** & \\
$\Lambda(1890)$ & $3/2^+$ & $\approx 100$ & **** & 0.8 &
  $\Sigma(2250)$ & $?^?$ & $\approx 100$ & ***  & \\
$\Lambda(2100)$ & $7/2^-$ & $\approx 200$ & **** & & & & & \\
$\Lambda(2110)$ & $5/2^+$ & $\approx 200$ & *** & & & & & \\
$\Lambda(2350)$ & $9/2^+$ & $\approx 150$ & *** & & & & & \\
\hline\hline
\end{tabular}
\label{tbl:hyperons}
\end{table*}

Unfortunately, for higher-mass resonances
(those in Table~\ref{tbl:hyperons} with a mass larger than 1.6 GeV),
there is not sufficient information to extract the necessary parameters,
especially their coupling constants to the cascade baryon.
The only available information relevant to the present reaction involving
the higher hyperon resonances are the $Y\to N \bar{K}$ partial decay
widths~\cite{PDG}, from which we can extract the magnitude of the
corresponding $NYK$ coupling constants.
They are displayed in Table~\ref{tbl:hyperons}.
Therefore, we consider only the diagrams (a)--(g) in Fig.~\ref{fig:diagram}
with $Y=Y'$ in (d), where the only additional parameter is the $\Xi YK$
coupling constant.
Also, in the following, we will restrict ourselves to spin-$1/2$ and -$3/2$
hyperons.
It then happens that, unless the $\Xi YK$ coupling constants are much
larger than the corresponding $NYK$ coupling constants, resonances with
$J^P=1/2^+$ and $3/2^-$ yield much smaller contributions to the reaction
amplitude as compared to the $J^P=1/2^-$ and $3/2^+$ resonance contributions.
This can be understood if we consider the limit of the intermediate hyperon
resonances being on the mass shell.
Then, the photoproduction amplitude becomes proportional to either the sum
of the baryon masses or their difference depending on the spin-parity
of the resonance, namely,
\begin{equation}
\begin{aligned}
M_{1/2^\pm} & \propto  (m_Y\mp m_N)(m_Y\mp m_\Xi)  ,  \\
M_{3/2^\pm} & \propto  (m_Y\pm m_N)(m_Y\pm m_\Xi)  ,
\end{aligned}\label{MMDF}
\end{equation}
where $M_{J^P}$ denotes the photoproduction amplitude involving the
intermediate hyperon with the spin-parity $J^P$.
Of course, this argument does not quite apply to low-mass resonances which
are far off-shell in the present reaction.
Among the $J^P=1/2^-$ and $3/2^+$ resonances, assuming the $\Xi YK$
coupling strength to be of the order of the $NYK$ coupling strength,
we find that only the $\Lambda(1800)1/2^-$ and $\Lambda(1890)3/2^+$
resonances contribute significantly.%
\footnote{Among the $\Sigma$ resonances, the only candidate is
$\Sigma(1750)$ with $J^P = 1/2^-$, which, however, has a very small
coupling constant $g_{N\Sigma K}^{}$.
For $\Lambda$ resonances, one can expect
$|g_{\Xi\Lambda K}^{}/g_{N\Lambda K}^{}| \le 1$ since
$g_{\Xi\Lambda K}^{} = g_{N\Lambda K}^{}$ for a singlet $\Lambda$ and
$g_{\Xi\Lambda K}^{}/g_{N\Lambda K}^{} = (1-4f)/(1+2f)$ for an octet
$\Lambda$~\cite{deS63}, which is less than 1 for $0<f<1$.}
We, therefore, consider only these two higher-mass resonances in the
present exploratory investigation instead of including all hyperon
resonances listed in Table~\ref{tbl:hyperons}.
As such, these two $\Lambda$ resonances may be viewed as representatives
of the spin-1/2 and -3/2 resonances, respectively, in the region of
1.8--2.0 GeV; they are employed here to indicate what features spin-1/2
and -3/2 resonances may introduce into the present reaction.

\begin{table*}[p]
\caption{Model parameters employed in the present calculation.
The last column cites the sources for the respective values, where PDG
refers to Ref.~\cite{PDG}.}
\begin{center}
\begin{tabular}{l@{\qquad}r@{\qquad}r}
\hline\hline
Nucleon        : & &  \\
$m_N^{} $     (MeV)                            & $938.3$    &  PDG \\
$\kappa^{}_{p\gamma},\  \kappa^{}_{n\gamma}$   & $1.79$, $-1.91$    & \\
\hline
$\Xi(1318)$    : & &  \\
$m_\Xi^{}$    (MeV)                          & $1318.0$    & \\
$\kappa_{\Xi^0\gamma}^{},\  \kappa^{}_{\Xi^-\gamma}$ & $-1.25$,  $0.35$ & PDG \\
\hline
$\Xi^*[=\Xi(1530)]$    : & &  \\
$m_{\Xi^*}^{} \ (\Gamma_{\Xi^*}^{}) $  (MeV) & $1533.0$ $(9.5)$ &  PDG \\
\hline\hline
$\Lambda(1116)$ : & &  \\
$m_\Lambda^{} $\  (MeV)                        & $1115.7$      &  PDG \\
$g_{N\Lambda K}^{}$                            & $-13.24$      &
SU(3) + ($f/d=0.575$ and $g_{NN\pi}=13.26$)\\
$g_{\Xi\Lambda K}^{}$                          & $3.52$         &
SU(3) + ($f/d=0.575$ and $g_{NN\pi}^{}=13.26$)\\
$g_{\Xi^*\Lambda K}^{}$                        & $5.58$         &
SU(3) + ($f_{N\Delta\pi}=2.23$)            \\
$g_{N\Lambda K^*}^{}\  (\kappa_{N\Lambda K^*}^{})$  & $-6.11$ $(2.43)$  &
Ref.~\cite{Nijmegen} (version NSC97f)              \\
$g_{\Xi\Lambda K^*}^{}\  (\kappa_{\Xi\Lambda K^*}^{})$ & $6.11$ $(0.65)$  &
Ref.~\cite{Nijmegen} (version NSC97f)              \\
$\kappa_{\Lambda\gamma}^{}$                & $-0.613$     &  PDG \\
\hline
$\Lambda(1405)$ : & &  \\
$m_\Lambda^{} \ (\Gamma_\Lambda^{}) $  (MeV)   & $1406.0$ $(50.0)$  & PDG \\
$g_{N\Lambda K}^{}$             & $\pm0.91$    &
SU(3) (flavor-singlet assumptions)      \\
$g_{\Xi\Lambda K}^{}$           & $\pm0.91$    &
SU(3) (flavor-singlet assumptions)      \\
$\kappa_{\Lambda\gamma}^{}$     &  $0.25$      &
Skyrme model \cite{Skyrme}, unitarized ChPT \cite{UChPT}        \\
\hline
$\Sigma(1193)$  : & &  \\
$m_\Sigma^{}$\ (MeV)             & $1193.0$         &  PDG \\
$g_{N\Sigma K}^{}$               & $3.58$         &
SU(3) + ($f/d=0.575$ and $g_{NN\pi}^{}=13.26$)\\
$g_{\Xi\Sigma K}^{}$             & $-13.26$    &
SU(3) + ($f/d=0.575$ and $g_{NN\pi}^{}=13.26$)\\
$g_{\Xi^*\Sigma K}^{}$           & $3.22$         &
 SU(3) + ($f_{N\Delta\pi}=2.23$)            \\
$g_{N\Sigma K^*}^{}\  (\kappa_{N\Sigma K^*}^{})$   & $-3.52$ $(-1.14)$  &
Ref.~\cite{Nijmegen} (version NSC97f)              \\
$g_{\Xi\Sigma K^*}^{}\  (\kappa_{\Xi\Sigma K^*}^{})$  & $-3.52$ $(4.22)$  &
Ref.~\cite{Nijmegen} (version NSC97f)              \\
$\kappa_{\Sigma^+\gamma}^{}$,  $\kappa_{\Sigma^0\gamma}^{}$,
$\kappa_{\Sigma^-\gamma}^{}$    & $1.46$, $0.65$, $-0.16$    &  PDG \\
\hline\hline
$\Lambda(1520)$ : & &  \\
$m_\Lambda^{} \ (\Gamma_\Lambda^{}) $  (MeV)   & $1519.5$ $(15.6)$  & PDG \\
$g_{N\Lambda K}^{}$                & $-10.90$  &
PDG, SU(3) (flavor-octet assumption)       \\
$g_{\Xi\Lambda K}^{}$              & $3.27$    &
PDG, SU(3) (flavor-octet assumption)       \\
$\kappa_{\Lambda\gamma}^{}$       &  $0.0$           & assumption  \\
\hline
$\Sigma(1385)$  : & &  \\
$m_\Sigma^{} \ (\Gamma_\Sigma^{}) $ (MeV)      & $1384.0$ $(37.0)$  &  PDG \\
$g_{N\Sigma K}^{}$                & $-3.22$      &
SU(3) + ($f_{N\Delta\pi}=2.23$)            \\
$g_{\Xi\Sigma K}^{}$             & $-3.22$          &
SU(3) + ($f_{N\Delta\pi}=2.23$)            \\
$f_{\Xi^*\Sigma K}^{}$           & $-2.83$        &
SU(3) + ($f_{\Delta\Delta\pi}=0.8$ from quark model) \\
$g^{(1)}_{N\Sigma K^*},\  g^{(2)}_{N\Sigma K^*}$  & $-5.47$, $0.0$ &
SU(3) + ($f_{N\Delta\rho}=5.5$)              \\
$g^{(1)}_{\Xi\Sigma K^*},\  g^{(2)}_{\Xi\Sigma K^*}$  & $-5.47$,
$0.0$   &
SU(3) + ($f_{N\Delta\rho}=5.5$)              \\
$\kappa_{\Sigma^+\gamma}^{}$,  $\kappa_{\Sigma^0\gamma}^{}$,
$\kappa_{\Sigma^-\gamma}^{}$   & $2.11$, $0.32$, $-1.47$    &
quark model \cite{Licht}             \\
\hline\hline
\end{tabular}
\label{tbl:cc}
\end{center}
\end{table*}
%

The interaction Lagrangians for constructing our model for the production
amplitudes shown in Fig.~\ref{fig:diagram} are given in Appendix.
The corresponding parameter values are in Tables~\ref{tbl:cc} and
\ref{tbl:trcc}.

\begin{table*}[t]
\begin{center}
\caption{Electromagnetic transition coupling constants employed in the
present calculation.
The last column cites the sources for the respective values.
PDG refers to Ref.~\cite{PDG}.}
\begin{tabular}{l@{\qquad}r@{\qquad}r}
\hline\hline
spin-1/2 $\leftrightarrow$ spin-1/2  transitions & $g_{BB'\gamma}$ & \\
\hline
$\Lambda(1116)\leftrightarrow\Lambda(1405)$      & $0.99$         &
quark model \cite{Darew}       \\
$\Lambda(1116)\leftrightarrow\Sigma(1193)$       & $1.61$         &
PDG + quark model               \\
$\Lambda(1405)\leftrightarrow\Sigma(1193)$       & $1.21$         &
quark model \cite{Darew}         \\
\hline\hline
spin-1/2 $\leftrightarrow$ spin-3/2 transitions   & $(g_1, g_2)$    & \\
\hline
$\Lambda(1116)\leftrightarrow\Lambda(1520)$      & ($\pm1.26$, $0.0$)&
CLAS data \cite{Taylor}              \\
$\Lambda(1116)\leftrightarrow\Sigma^0(1385)$     & ($2.81$,$-2.37$)    &
chiral quark model \cite{Wagner}     \\
$\Sigma^0(1193)\leftrightarrow\Lambda(1520)$     & ($\pm2.22$, $0.0$)&
PDG based on SU(3) assumption                    \\
$\Sigma^+(1193)\leftrightarrow\Sigma^+(1385)$    & ($2.68$, $-0.72$)   &
 chiral quark model  \cite{Wagner}    \\
$\Sigma^0(1193)\leftrightarrow\Sigma^0(1385)$    & ($0.40$, $ 0.31$)   &
 chiral quark model  \cite{Wagner}    \\
$\Sigma^-(1193)\leftrightarrow\Sigma^-(1385)$    & ($1.15$, $-0.58$)   &
 chiral quark model  \cite{Wagner}    \\
$\Xi^0(1530)\leftrightarrow\Xi^0(1318)$     & ($3.02$, $-2.40$)   &
chiral quark model \cite{Wagner}     \\
$\Xi^-(1530)\leftrightarrow\Xi^-(1318)$    & ($0.56$, $-0.16$)   &
chiral quark model \cite{Wagner}     \\
\hline\hline
$g^c_{KK^*\gamma}$                   &   $0.41$          & PDG + SU(3) \\
$g^0_{KK^*\gamma}$                   &  $-0.63$       & PDG + SU(3)
\\
\hline\hline
\end{tabular}
\label{tbl:trcc}
\end{center}
\vspace{-3ex}
\end{table*}
%

Before presenting our results, we mention that the free parameters in the
present model calculation include:
\begin{itemize}

\item[a)]
The pseudoscalar-pseudovector (ps-pv) mixing parameter $\lambda$ in the
meson-baryon ($BYK$) vertex for spin-1/2 baryons $B$ and $Y$ in
Eq.~(\ref{BYK12}).
Note that, in principle, because of the Goldstone-boson nature of kaons,
chiral symmetry demands the pseudovector-coupling ($\lambda=0$) choice
at least for energies near the threshold.
(Strictly speaking, of course, chiral symmetry holds only in the massless
kaon limit.)
Nevertheless, here we consider both the extreme possibilities: $\lambda=0$ and
$\lambda=1$.

\item[b)]
The signs of the hadronic and electromagnetic transition coupling
constants, $g_{B\Lambda K}^{}=\pm 0.91$ for $\Lambda(1405)$
(Table~\ref{tbl:cc}), and
$g^{}_{\Lambda\Lambda'\gamma}=\pm 1.26$ and
$g^{}_{\Sigma\Lambda'\gamma}=\pm 2.22$ for the transitions
$\Lambda(1116) \leftrightarrow \Lambda(1520)$ and
$\Sigma(1193) \leftrightarrow \Lambda(1520)$, respectively
(Table~\ref{tbl:trcc}).
The phases of those coupling constants are not uniquely fixed yet.%
\footnote{For those diagrams in Fig.~\ref{fig:diagram} containing the
$NYK$ and $\Xi Y'K$ vertices with $Y=Y'$, their corresponding total phases
can be fixed by SU(3) flavor symmetry.}

\item[c)]
The cutoff parameter $\Lambda_B$ and the exponent $n$ in the baryonic form
factor in Eq.~(\ref{eq:ffB}).
We take these parameters to be the same for all the intermediate baryons.

\item[d)]
The product of the coupling constants, $g_{N\Lambda K}^{}g_{\Xi\Lambda K}^{}$,
for higher-mass resonances, $\Lambda(1800)1/2^-$ and $\Lambda(1890)3/2^+$,
as has been explained before.

\end{itemize}

\section{Results}

We now turn our attention to the results of the present model.
The strategy we follow here is as follows.
For a given choice of the ps-pv mixing parameter and the signs of the
coupling constants mentioned in items a) and b) above, we adjust the
parameters of the baryonic form factor mentioned in item c) to reproduce
the preliminary total cross section data~\cite{Guo} in the
$\gamma p \to K^+K^+\Xi^-$ channel.
We note that, in general, short-range processes are very sensitive to the
hadronic off-shell form factors.
In the present reaction, not only the absolute normalization, but also the
energy dependence of the total cross section is found to be extremely
sensitive to the baryonic form factor given in Eq.~(\ref{eq:ffB}) for
low-mass resonance contributions.
We also mention that the coupling constants used in the present model
calculation were taken as the centroid values of those quoted in
PDG~\cite{PDG} and other hadron model predictions whenever available.
Hence, since these coupling constants generally are subject to considerable
uncertainties, large uncertainties of comparable size can be expected for the
resulting fitted parameters of our model.

\subsection{Low-mass resonances}\label{sec:lowmass}

\begin{figure*}[t!]\centering
\includegraphics[width=0.9\textwidth,angle=0,clip]{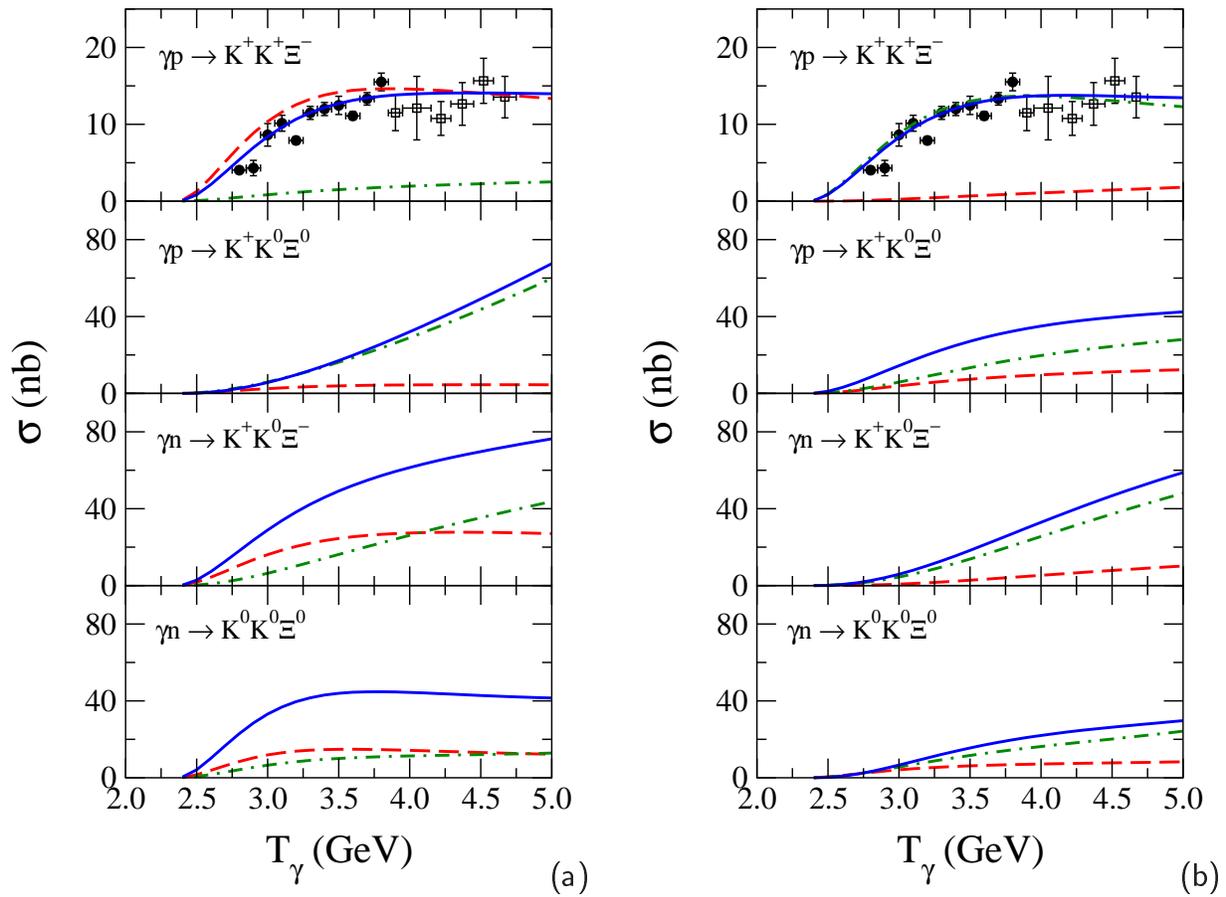}
\caption{\label{fig:txsc_lmR}%
(Color online) Total cross sections for $\gamma N\to K K \Xi$ according to the
mechanisms shown in Fig.~\ref{fig:diagram} as a function of photon incident
energy $T_\gamma$ for (a) pseudovector ($\lambda=0$) and (b) pseudoscalar
($\lambda=1$) couplings.
The dashed curves correspond to the contribution from the diagrams involving
only the spin-1/2 hyperons, while the dash-dotted curves correspond to the
contribution from the diagrams involving one or more spin-3/2 hyperons.
The solid curves represent the total contribution.
The (preliminary) data are from Ref.~\cite{Guo}, where the open boxes are
obtained without the differential cross section measurement.}
\end{figure*}
%

We will first discuss the results considering only the low-mass hyperons
[$\Lambda(1116)$, $\Lambda(1405)$, $\Lambda(1520)$, $\Sigma(1190)$,
$\Sigma(1385)$, $\Xi(1318)$, and $\Xi(1530)$]
whose relevant coupling constants can be determined from independent sources
as given in Tables~\ref{tbl:cc} and \ref{tbl:trcc}.%
\footnote{We have also considered the $\Delta(1232)$ resonance which may
contribute to the present reaction through diagram (c) in
Fig.~\ref{fig:diagram}. However, its contribution is negligibly small.}
Here, the signs of the coupling constants mentioned in item b) in the
previous Section are chosen to be all positive.
The adjusted baryonic form-factor parameters are $\Lambda_B=1225$ MeV and
$n=2$ for both the ps- ($\lambda=1$) and pv-coupling ($\lambda=0$)
choices.%
\footnote{We have varied the value of $n$ together with the cutoff parameter
$\Lambda_B$, and found that $n=2$ gives a good description of the
preliminary total cross section data not only in the magnitude but in
the energy dependence as well.
For $n=1$, the total cross section keeps increasing as the incident photon
energy increases.}
The resulting total cross sections are shown in Fig.~\ref{fig:txsc_lmR} for
both choices of the ps-pv mixing parameter $\lambda =0$ and $\lambda=1$.
The solid curves in Fig.~\ref{fig:txsc_lmR} are the sum of all
contributions and reproduce the basic features exhibited by the available
preliminary data from the CLAS Collaboration~\cite{Guo} quite well.
The predicted total cross sections for the other channels
$\gamma p \to K^+ K^0 \Xi^0$, $\gamma n \to K^+ K^0 \Xi^-$, and
$\gamma n \to K^0 K^0 \Xi^0$ are much larger than those for the
$\gamma p \to K^+ K^+ \Xi^-$ channel.
Both choices of the ps-pv mixing parameter $\lambda (=0,1)$ yield results
(solid curves) which are close to each other in the energy region considered,
especially, in the $\gamma p \to K^+ K^+ \Xi^-$ channel where the data exist.
However, their dynamical contents are quite different.
The dashed curves correspond to the contributions from the diagrams in
Fig.~\ref{fig:diagram} which involve only the spin-1/2 hyperons in the
intermediate state, while the dash-dotted curves correspond to those
involving \emph{one or more\/} spin-3/2 hyperons.
One can see that the latter dominates in all the channels for the
ps-coupling choice, while the former can be the dominant contribution for the
pv-coupling choice depending on the reaction channel.
Note, especially, that in the $\gamma p \to K^+ K^+ \Xi^-$ channel, the
dominant contribution comes from the diagrams involving only spin-1/2
hyperons for the pv-coupling and from the diagrams involving one or more
spin-3/2 hyperons in the case of the ps-coupling choice.

\begin{figure*}[t!]\centering
\includegraphics[width=0.9\textwidth,angle=0,clip]{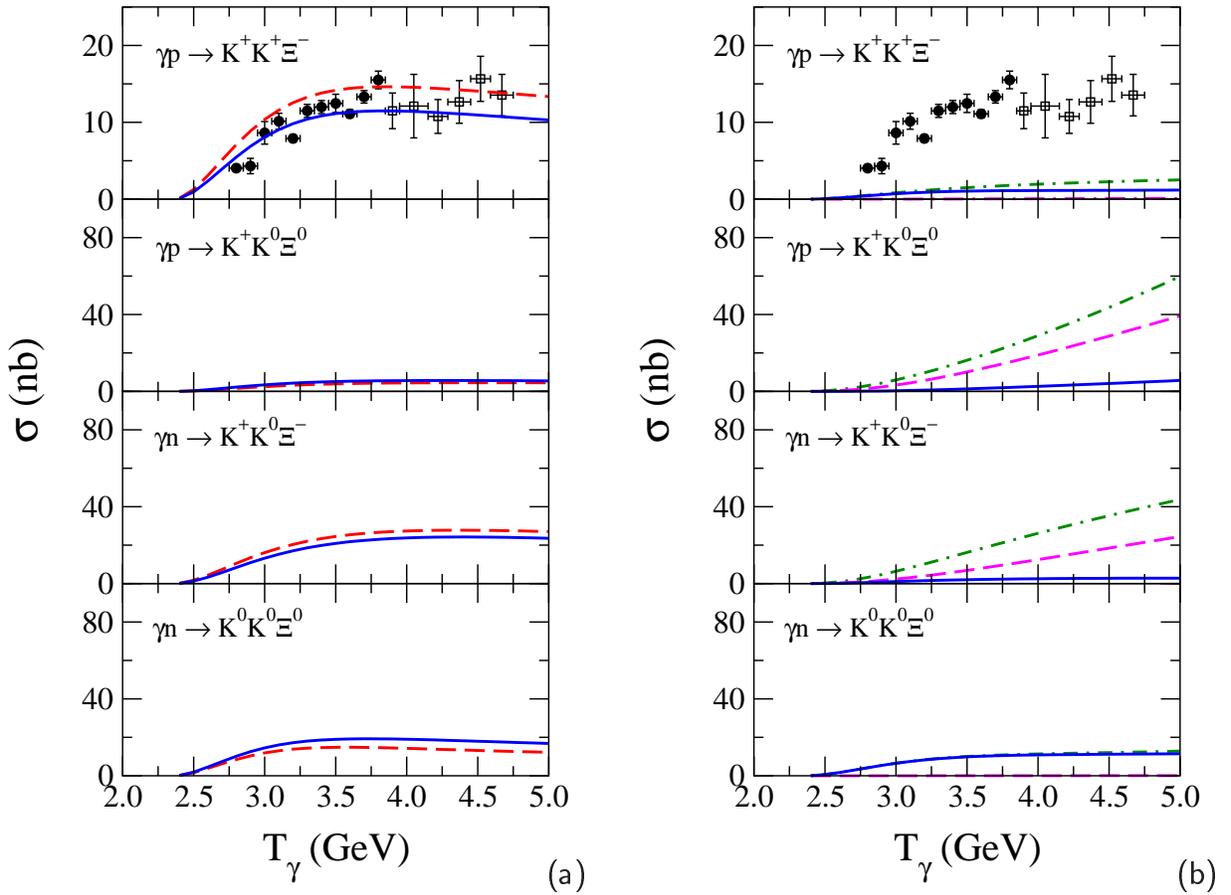}
\caption{\label{fig:txsc_lmR_pv_Y32Y12}%
(Color online)
(a) Breakdown of the spin-1/2 hyperon contributions to the total cross
section for $\gamma N\to K K \Xi$ shown in Fig.~\ref{fig:txsc_lmR}(a) for
the pv-coupling choice according to the mechanisms shown in
Fig.~\ref{fig:diagram}.
The solid curves correspond to the contributions from the radiative
transition diagram Fig.~\ref{fig:diagram}(d) with $Y \ne Y'$ and the dashed
curves to the total contribution from the spin-1/2 hyperons [same as the
dashed curves in Fig.~\ref{fig:txsc_lmR}(a)].
\ (b) Same as (a) but for the processes involving one or more spin-3/2
hyperon.
The dashed curves correspond to the spin-3/2 resonance contribution
Fig.~\ref{fig:diagram}(d) with $Y=Y'$ and the dash-dotted curves to the
total contribution [same as the dash-dotted curves in
Fig.~\ref{fig:txsc_lmR}(a)].
The solid curves correspond to the spin-3/2 $\leftrightarrow$ spin-1/2
radiative transition diagram Fig.~\ref{fig:diagram}(d) with $Y\ne Y'$.
Contributions from the other diagrams in Fig.~\ref{fig:diagram} are too
small and, therefore, not shown here.
The (preliminary) data are from Ref.~\cite{Guo}.}
\end{figure*}

Shown in Fig.~\ref{fig:txsc_lmR_pv_Y32Y12} is the dynamical content of our
model for the pv-coupling choice. We see that among the spin-1/2 hyperon
contributions [Fig.~\ref{fig:txsc_lmR_pv_Y32Y12}(a)] to the total cross section
results in Fig.~\ref{fig:txsc_lmR}, the dominant one is by far the spin-1/2
$\leftrightarrow$ spin-1/2 radiative transition represented by the diagram
Fig.~\ref{fig:diagram}(d) with $Y\ne Y'$. The other diagrams yield negligible
contributions. Among the diagrams containing one or more spin-3/2 hyperon, the
dominant contributions are either the spin-3/2 $\leftrightarrow$ spin-1/2
radiative transition or the spin-3/2 resonance contribution,
Fig.~\ref{fig:diagram}(d) with $Y=Y'$, depending on the reaction channel.

%

\begin{figure}[t!]
\includegraphics[width=0.4\textwidth,angle=0,clip]{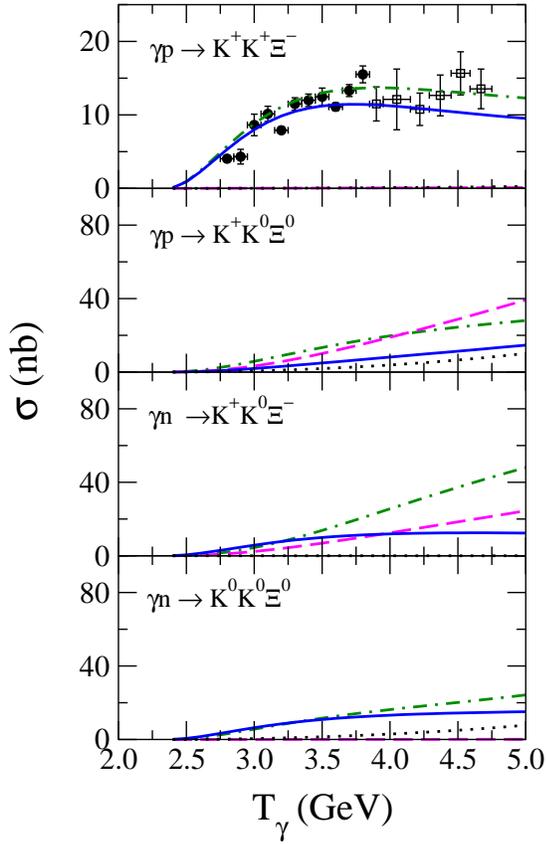}
\caption{\label{fig:txsc_lmR_ps_Y32}%
(Color online)
Same as Fig.~\ref{fig:txsc_lmR_pv_Y32Y12}(b) but with the ps-coupling choice.
The solid curves correspond to the contribution from the radiative
transition diagram Fig.~\ref{fig:diagram}(d) with $Y\ne Y'$ and the dashed
curves to the spin-3/2 resonance contribution Fig.~\ref{fig:diagram}(d)
with $Y=Y'$.
The dotted curves represent the $\Xi(1530) \leftrightarrow \Xi(1318)$
radiative transition of Fig.~\ref{fig:diagram}(e).
The dash-dotted curves are the total spin-3/2 contributions as shown in
Fig.~\ref{fig:txsc_lmR}(b).
The other diagrams of Fig.~\ref{fig:diagram} yield too small contributions
and are not shown.}
\end{figure}

Similarly, Fig.~\ref{fig:txsc_lmR_ps_Y32} shows the dynamical content of our
model for the ps-coupling choice.
Here only those diagrams involving one or more spin-3/2 hyperon are shown.
The contributions from the spin-1/2 hyperons are small in all the reaction
channels as shown in Fig.~\ref{fig:txsc_lmR}.
As one can see, in the $\gamma p\to K^+ K^+ \Xi^-$ and
$\gamma n\to K^0 K^0 \Xi^0$ channels, the spin-3/2 $\leftrightarrow$
spin-1/2 radiative transitions (solid curves) are, by far, the dominant
processes.
In other channels, there are competing mechanisms, namely, the spin-3/2
resonance contribution represented by Fig.~\ref{fig:diagram}(d) with $Y=Y'$
(dashed curves) and the $\Xi(1530) \leftrightarrow \Xi(1318)$ radiative
transition, Fig.~\ref{fig:diagram}(e) (dash-dotted curves), in addition to
the spin-3/2 $\leftrightarrow$ spin-1/2 radiative transitions.

%

\begin{figure}[t!]\centering
\includegraphics[width=0.85\columnwidth,angle=0,clip]{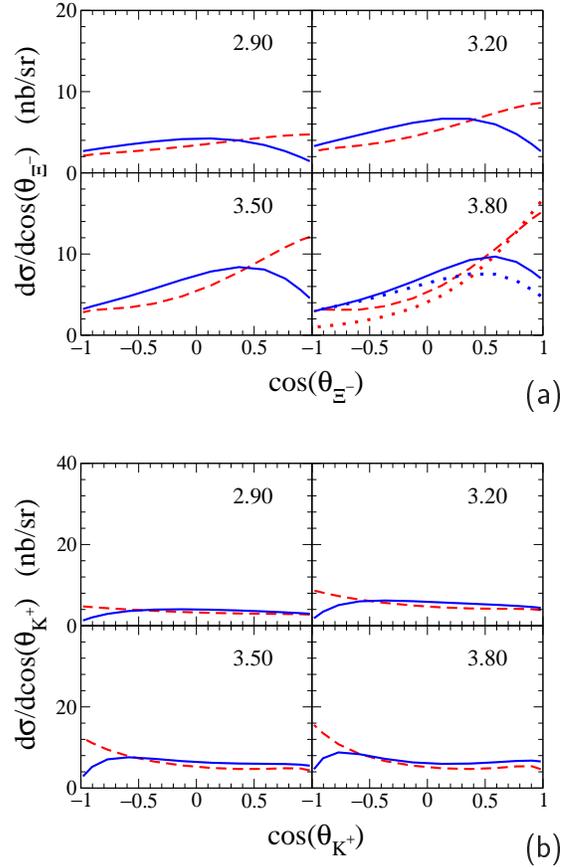}
\caption{\label{fig:dxsc_lmR}%
(Color online)
Predicted angular distributions for the (a) $\Xi^-$ and (b)
$K^+$ particles for $\gamma p\to K^+ K^+ \Xi^-$ in the center-of-mass frame
corresponding to the results of Fig.~\ref{fig:txsc_lmR} with the pv- (solid
curves) and ps-coupling (dashed curves) choices.
The number in the right upper corner of each figure indicates the incident
photon energy $T_\gamma$ in units of GeV.
The dotted curves at $T_\gamma =3.80$ GeV in (a) represent the radiative
transition contributions corresponding to Fig.~\ref{fig:diagram}(d)
with $Y\ne Y'$.}
\end{figure}

Figure~\ref{fig:dxsc_lmR} shows the model predictions for the angular
distributions of the produced cascade $\Xi^-$ and kaon $K^+$ in the
center-of-mass frame of the total system.
They correspond to the total cross section results of
Fig.~\ref{fig:txsc_lmR}.
We present the results for four different photon energies, spanning the
energy region relevant to ongoing cascade photoproduction experiment at
JLab~\cite{Guo,Guo1}.
One can see that the cascade angular distribution becomes forward-angle
peaked as the energy increases, while the $K^+$ distribution becomes
backward-angle peaked.
This tendency is stronger with the ps-coupling choice.
Here, we note that the shapes of the angular distributions are sensitive to
the production mechanism.
In particular, the exhibited shapes of the angular distributions are due to
the dominance of the spin-1/2 $\leftrightarrow$ spin-1/2 hyperon radiative
transitions [diagram Fig.~\ref{fig:diagram}(d)] for the pv-coupling choice and
of the spin-3/2 $\leftrightarrow$ spin-1/2 hyperon radiative transitions for
the ps-coupling choice, as shown by the corresponding dotted lines in the left
panel of Fig.~\ref{fig:dxsc_lmR} for $T_\gamma=3.80$ GeV.
Note that the different behavior of the angular distribution in the forward
(backward) $\Xi^-$ ($K^+$) angles between the pv- and ps-coupling choices is
caused by the radiative transition diagrams.
We also note that the $\Xi^-$ and $K^+$ angular distributions are not
completely independent of each other, because, in the center-of-momentum
(\mbox{c.m.}) frame of the system, energy-momentum conservation demands that
if one of the $K^+$ is emitted in forward (backward) angles, the $\Xi^-$ be
emitted necessarily in backward (forward) angles for sufficiently high
incident energies.
We emphasize that the shape of the angular distribution can change completely
from that predicted in Fig.~\ref{fig:dxsc_lmR} if the dominant production
mechanism is the $t$-channel $K$-exchange current
[diagrams Figs.~\ref{fig:diagram}(a,b)] instead of the radiative
transition current.%
\footnote{Among the various production mechanisms considered
in this work, only the $t$-channel $K$-exchange process
[diagram Fig.~\ref{fig:diagram}(a)] exhibits the backward-peaked angular
distribution for $\Xi^-$.}
As we shall show in the following subsection, such a situation is possible
when contributions from the higher mass resonances are considered.

%

\begin{figure}[t!]\centering
\includegraphics[width=0.85\columnwidth,angle=0,clip]{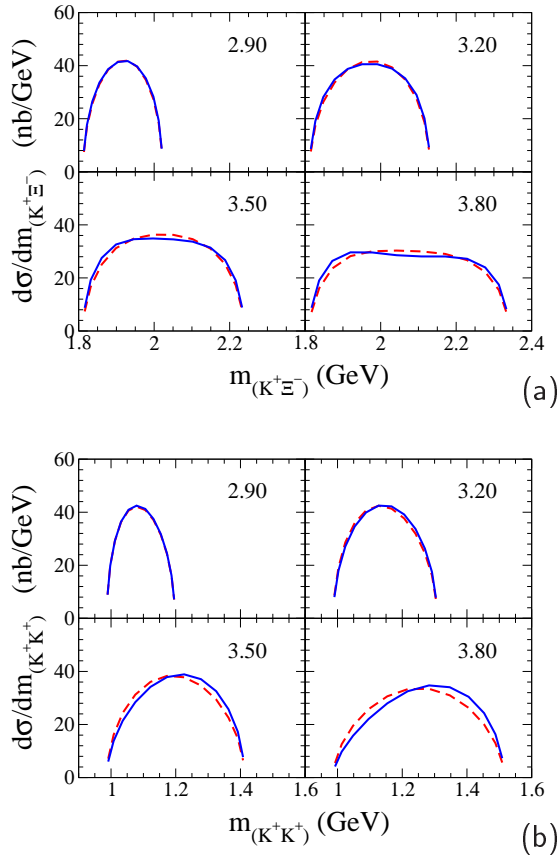}
\caption{\label{fig:dxsc_invmass_lmR}%
(Color online)
Predicted invariant-mass distributions for (a) $K^+\Xi^-$ and
(b) $K^+K^+$ for $\gamma p\to K^+ K^+ \Xi^-$ corresponding to the results of
Fig.~\ref{fig:txsc_lmR} with the pv- (solid curves) and ps-coupling (dashed
curves) choices.
The number in the right upper corner of each subpanel indicates the
incident photon energy in units of GeV.}
\end{figure}

Displayed in Fig.~\ref{fig:dxsc_invmass_lmR} are the predictions for the
$K^+\Xi^-$ and $K^+K^+$ invariant-mass distributions, respectively, in $\gamma
p \to K^+ K^+ \Xi^-$.
Again, they correspond to the total cross-section results in
Fig.~\ref{fig:txsc_lmR}.
As has been pointed out already in connection with the total cross sections
in Fig.~\ref{fig:txsc_lmR}, here the results do not exhibit any resonance
structure in the $K^+\Xi^-$ invariant mass distribution because all the
hyperon resonances considered here lie below the production threshold.
The absence of any structure in the $K^+K^+$ invariant mass distribution is
due to the absence of $S=+2$ exotic meson production.

%

\subsection{Higher-mass resonances (I)}\label{sec:highmass}

We now explore the possible influence of the higher-mass $S=-1$ hyperon
resonances.
Since the higher hyperon resonances lie close to or above the threshold
energy of $\Xi$ production (see Table~\ref{tbl:hyperons}), it is natural
to expect them to play a more prominent role than the low-mass hyperons.
However, because of the lack of sufficient information to extract the
necessary parameter values associated with these high-mass resonances,
here we consider the hyperon resonances $\Lambda(1800)1/2^-$ and
$\Lambda(1890)3/2^+$, as discussed in Sec.~II.
Since no parameters other than the $NYK$ coupling constants
$g_{NYK}^{}$ are known, we consider the diagrams of
Figs.~\ref{fig:diagram}(a)--(g), where $Y=Y'$ in diagram
Fig.~\ref{fig:diagram}(d).
This introduces the coupling constant $g_{\Xi YK}^{}$ as the additional
free parameter as we neglect the magnetic moments of these $\Lambda$
resonances.
Then, since only the product $g_{NYK}^{}g_{\Xi YK}^{}$ enters into the
calculation of these processes, we explore the influence of the higher-mass
resonances as a function of this product of the coupling constants.

\begin{figure*}[t!]\centering
\includegraphics[width=0.9\textwidth,angle=0,clip]{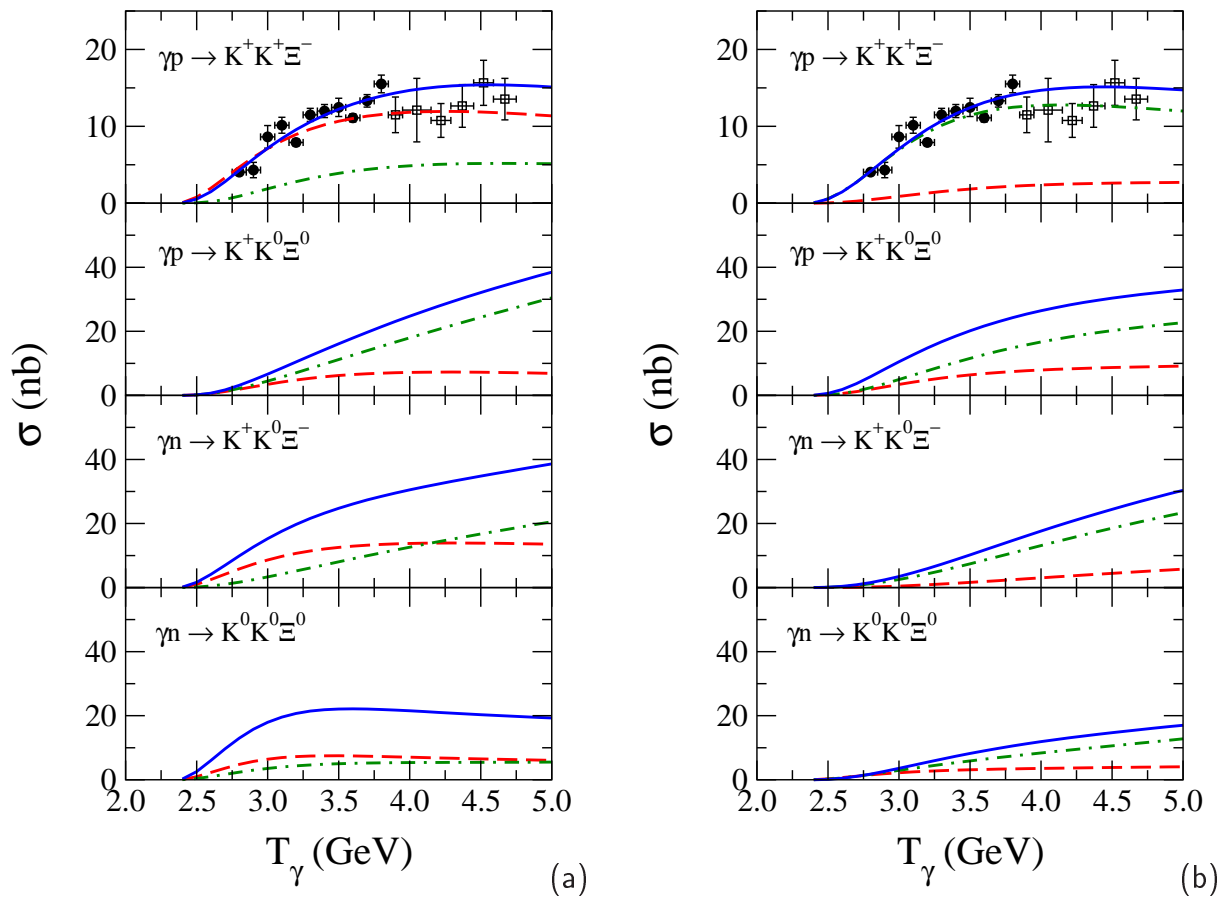}
\caption{\label{fig:txsc_hmR}%
(Color online)
Same as Fig.~\ref{fig:txsc_lmR}, but with the addition of the
higher-mass hyperons $\Lambda(1800)1/2^-$ and $\Lambda(1890)3/2^+$.}
\end{figure*}

Figure~\ref{fig:txsc_hmR} shows the results for the total cross sections
for $\gamma N \to K K \Xi$ when the above-mentioned higher-mass resonances
are included in addition to the low-mass resonances discussed in the previous
subsection.
Here, the product of the coupling constants is fixed to be
$g_{N\Lambda K}^{}g_{\Xi\Lambda K}^{} = 2$ for both resonances assuming that
$g_{\Xi\Lambda K}^{}$ has the same order of magnitude as $g_{N\Lambda K}^{}$.
The values of $\Lambda_B=1170$ MeV and $n=2$ were readjusted to reproduce
the measured total cross section~\cite{Guo}.
Everything else is kept the same as in the previous subsection.
We then obtain essentially the same results as in the previous section
where only the low-mass hyperons were considered.
This shows that the total cross sections alone are unable to distinguish
the contributions from the low- and high-mass hyperons.
We emphasize, however, that this is not the case when we consider the
total cross section in conjunction with the angular
distributions of the produced cascade and/or kaons.
In fact, for example, when we insist that the resulting shape of the
$\Xi^-$ and/or $K^+$ angular distributions be of the form shown in
Fig.~\ref{fig:dxsc_hmR} (see below), we could not reproduce the measured
energy dependence of the total cross section in $\gamma p \to K^+ K^+ \Xi^-$
without both the high-mass hyperons considered here.
This reveals that this reaction is, in fact, suited for extracting
information on the high-mass hyperons.
We also repeat that the $t$-channel $\bar{K}N \to K\Xi$ process and
meson-production processes are absent in the present reaction, which is
a feature that makes this reaction more suitable for studying high-mass
hyperon resonances.

%

\begin{figure*}[t!]
\includegraphics[width=0.9\textwidth,angle=0,clip]{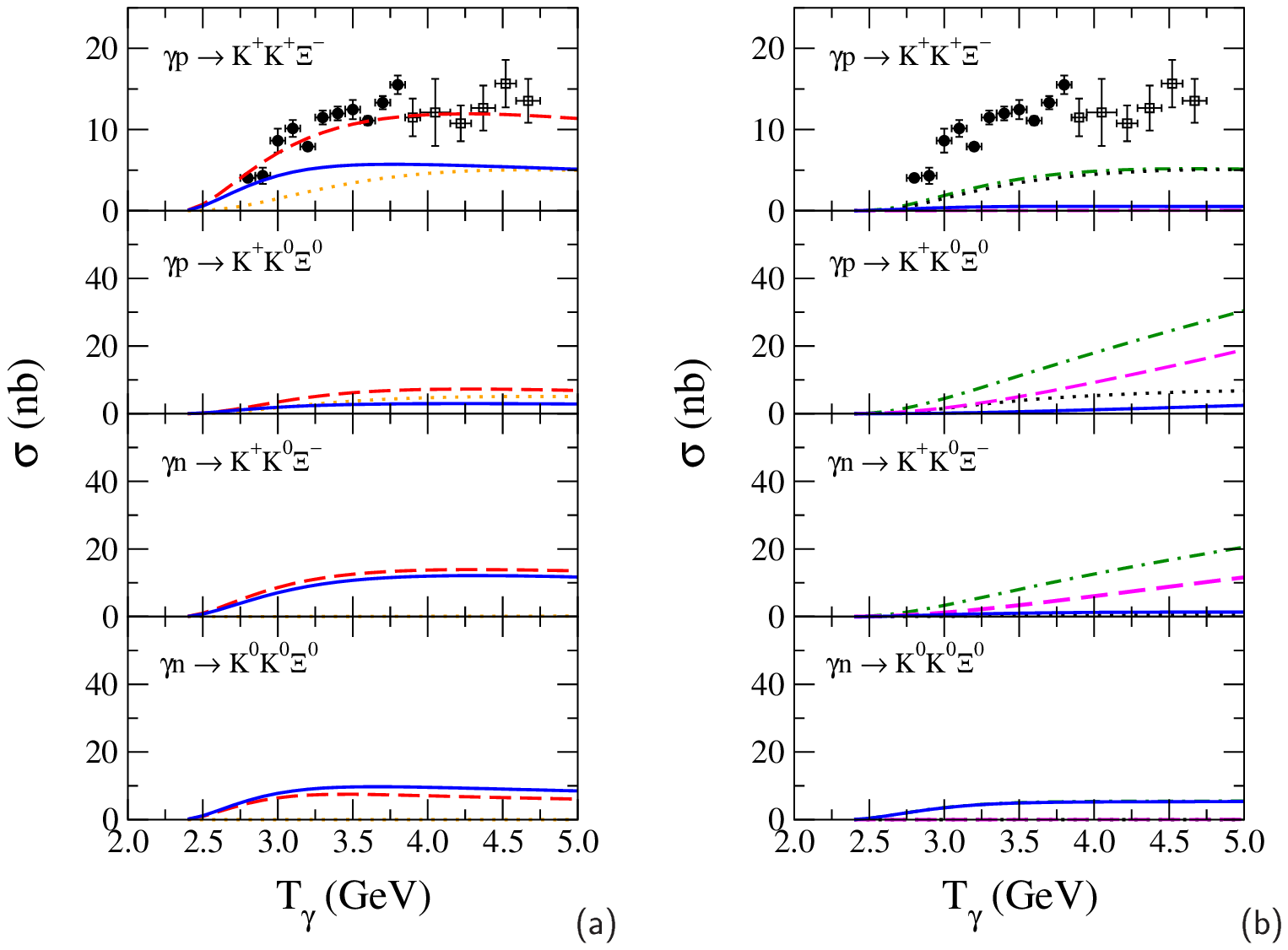}
\caption{\label{fig:txsc_hmR_pv_Y32Y12}%
(Color online)
Same as Fig.~\ref{fig:txsc_lmR_pv_Y32Y12}, but for the results shown in
Fig.~\ref{fig:txsc_hmR}(a) with the pv-coupling choice.
\ (a) The solid curves correspond to the contribution from the
radiative-transition diagram Fig.~\ref{fig:diagram}(d) with $Y\ne Y'$ and
the dotted curves to the $K$-exchange diagrams of
Figs.~\ref{fig:diagram}(a,b).
The dashed curves represent the total contributions from the spin-1/2
hyperons and are taken over from Fig.~\ref{fig:txsc_hmR}.
\ (b) The dashed curves correspond to the diagram Fig.~\ref{fig:diagram}(d)
with $Y=Y'$ and the dotted curves to the $K$-exchange diagrams of
Figs.~\ref{fig:diagram}(a,b).
The solid curves are for the spin-3/2 $\leftrightarrow$ spin-1/2 radiative
transition diagram Fig.~\ref{fig:diagram}(d) with $Y\ne Y'$.
The dash-dotted curves are the total contribution taken over from
Fig.~\ref{fig:txsc_hmR}(a).
The other contributions are too small and are not shown here.}
\end{figure*}

The dynamical content of the model considered in this subsection is shown in
Fig.~\ref{fig:txsc_hmR_pv_Y32Y12} for the case of the pv-coupling choice.
As can be seen, contrary to the results of the previous subsection, we now
have the $t$-channel $K$-exchange mechanism [Figs.~\ref{fig:diagram}(a,b)]
competing with the radiative transition mechanism depending on the reaction
channels.
Similar observations can be made for the case of the ps-coupling choice.
Note that in the $\gamma p \to K^+ K^+ \Xi^-$ channel, the $K$-exchange is,
by far, the dominant contribution arising from the diagrams involving the
spin-3/2 hyperons.
Of course, in the $\gamma n \to K^0 K^0 \Xi^0$ channel, the $K$-exchange
mechanism is simply absent.

%

In Fig.~\ref{fig:dxsc_hmR}, we show the model predictions for the angular
distributions of the produced $\Xi^-$ and $K^+$ in the \mbox{c.m.} frame
of the total system.
They correspond to the total cross section results of Fig.~\ref{fig:txsc_hmR}.
Here we see that the shape of the angular distributions are just the opposite
to those shown in Fig.~\ref{fig:dxsc_lmR}, where only the low-mass hyperons
were considered.
Here, the backward- (forward-) peaked angular distribution of $\Xi^-$
($K^+$) is due to the dominance of the $t$-channel $K$-exchange process,
Figs.~\ref{fig:diagram}(a,b).
This is illustrated in Fig.~\ref{fig:dxsc_hmRa} at $T_\gamma = 3.80$ GeV for
the pv- and ps-coupling choices.
It is a simple matter of kinematics that the $t$-channel processes contribute
mostly for low $t$ and high incident energies which leads to the
forward-peaked $K^+$ angular distribution as the incident photon energy
increases.
In the \mbox{c.m.} frame this, in turn, leads to the backward-peaked
$\Xi^-$ angular distribution.
It is obvious, then, that the angular distributions can tell us about the
$\Xi^-$ production mechanism, in particular, whether the dominant mechanism
is the radiative transitions (as in the previous subsection) or the
$t$-channel $K$-exchange (as shown here).
In the $\gamma n \to K^0 K^0 \Xi^0$ reaction channel, the latter mechanism,
of course, is absent.

\begin{figure}[t!]\centering
\includegraphics[width=0.85\columnwidth,angle=0,clip]{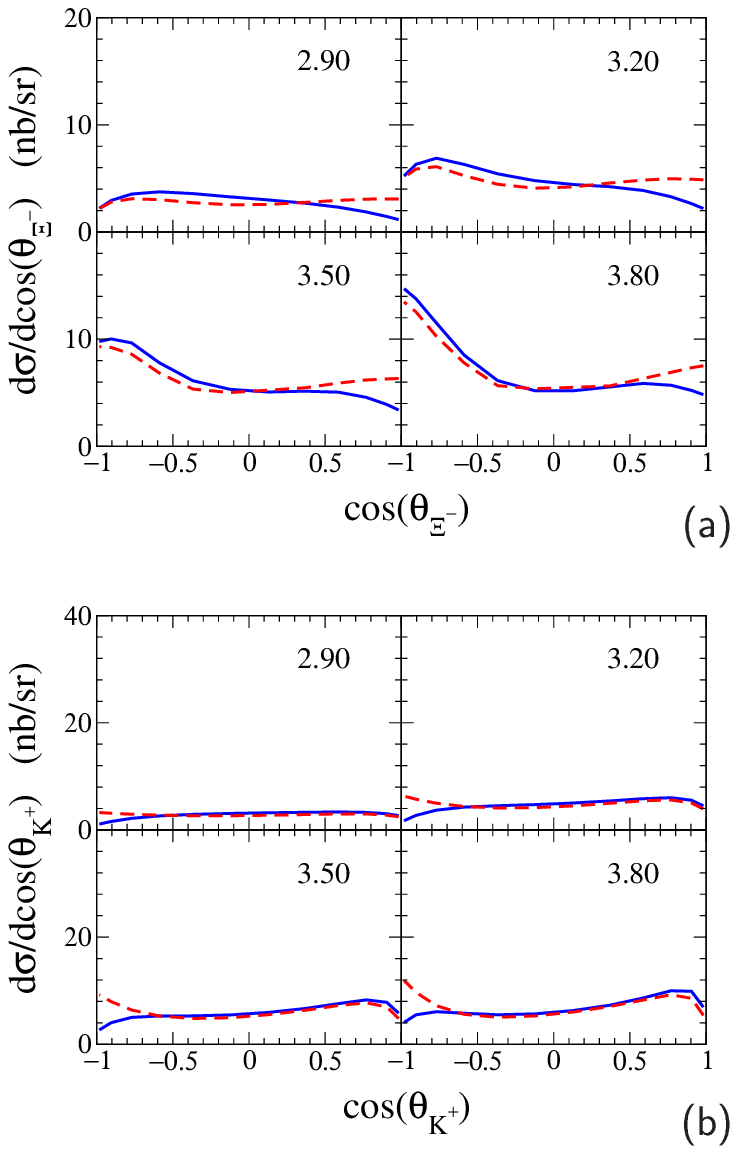}
\caption{\label{fig:dxsc_hmR}%
(Color online)
Same as Fig.~\ref{fig:dxsc_lmR}, but for the results of
Fig.~\ref{fig:txsc_hmR}.}
\label{fig9}
\end{figure}
%

\begin{figure}[b!]\centering
\includegraphics[width=0.95\columnwidth,angle=0,clip]{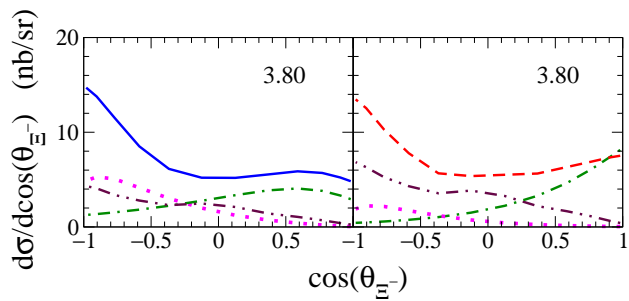}
\caption{\label{fig:dxsc_hmRa}%
(Color online)
Same as Fig.~\ref{fig9} at $T_\gamma = 3.80$ GeV.
In the left panel, the dotted and dash-dotted
curves correspond to the contributions from the $t$-channel $K$-exchange
and radiative transition processes, respectively,
[Figs.~\ref{fig:diagram}(a,b,d) with $Y\ne Y'$] involving only the spin-1/2
hyperons.
The dash--double-dotted curve is due to the $t$-channel $K$-exchange
involving one or more spin-3/2 hyperons.
They all correspond to the pv-coupling choice.
Same in the right panel, but with the ps-coupling choice.}
\end{figure}
%

\begin{figure}[t!]\centering
\includegraphics[width=0.85\columnwidth,angle=0,clip]{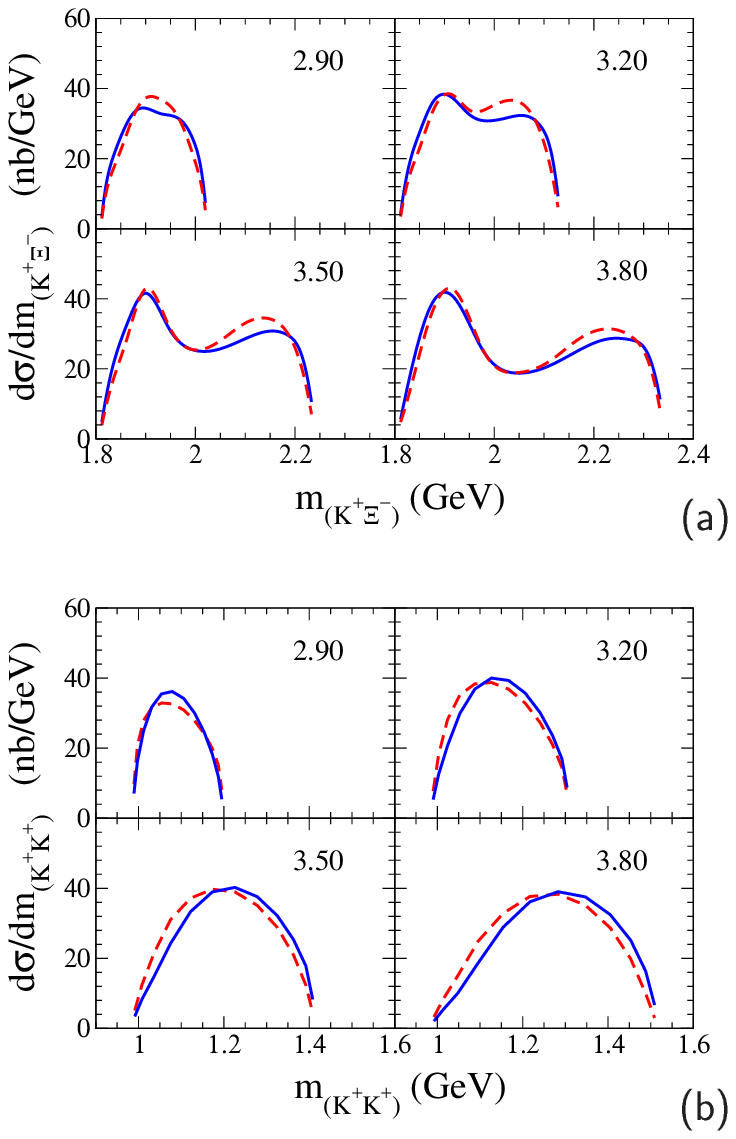}
\caption{\label{fig:dxsc_invmass_hmR}%
(Color online)
Same as Fig.~\ref{fig:dxsc_invmass_lmR}, but for the results of
Fig.~\ref{fig:txsc_hmR}.}
\end{figure}
%

In Fig.~\ref{fig:dxsc_invmass_hmR}, we display the predictions for the
$K^+\Xi^-$ and $K^+K^+$ invariant-mass distributions in
$\gamma p \to K^+ K^+ \Xi^-$.
Again, they correspond to the total cross section results in
Fig.~\ref{fig:txsc_hmR}.
Since we do not have $S=+2$ exotic meson production, the $K^+K^+$
invariant-mass distributions are very similar to those shown in
Fig.~\ref{fig:dxsc_invmass_lmR} and have no structure.
However, the $K^+\Xi^-$ invariant-mass distributions are quite different
from those shown in Fig.~\ref{fig:dxsc_invmass_lmR} for, here, they exhibit
two bump structures as the incident photon energy increases.
The $\Lambda(1800)$ hyperon is just below the threshold and contributes to
the sharp rise of the $K^+\Xi^-$ invariant-mass distribution near the
threshold.
The bump at lower invariant mass is due to the $\Lambda(1890)$ hyperon.
The second bump at higher invariant mass is due to the same hyperon
$\Lambda(1890)$, but comes from the diagram Fig.~\ref{fig:diagram}(b).
We emphasize that the appearance of a second bump at a higher invariant mass
is a general feature of two-meson production reactions and should not be
confused with the existence of another resonance with a higher mass.
Note that the position of the second bump changes depending on the photon
energy, which evidently makes it clear that the structure does not
come from a resonance.

As we have shown, the angular distributions of $\Xi^-$ and $K^+$ are
sensitive to the production mechanism of $\gamma p \to K^+ K^+ \Xi^-$.
In particular, depending on whether the dominant mechanism is the radiative
transition or $t$-channel $K$-exchange, the shape of the angular
distribution can change completely.
However, one should keep in mind that most of the parameters in the present
work have been fixed based on quark models and/or SU(3) symmetry
considerations in combination with independent experimental information
whenever available.
The parameter values estimated in this way may, therefore, be subject to
considerable uncertainties.
In particular, one cannot completely discard the possibility that reaction
mechanisms other than the $t$-channel $K$-exchange might lead,
for example, to backward- (forward-) peaked $\Xi^-$ ($K^+$) angular
distribution through the interference with the radiative transition
mechanisms.
It is interesting, therefore, to look for an independent observable which
is also sensitive to the production mechanisms other than the
differential cross sections.

\begin{figure}[t!]\centering
\includegraphics[width=0.85\columnwidth,angle=0,clip]{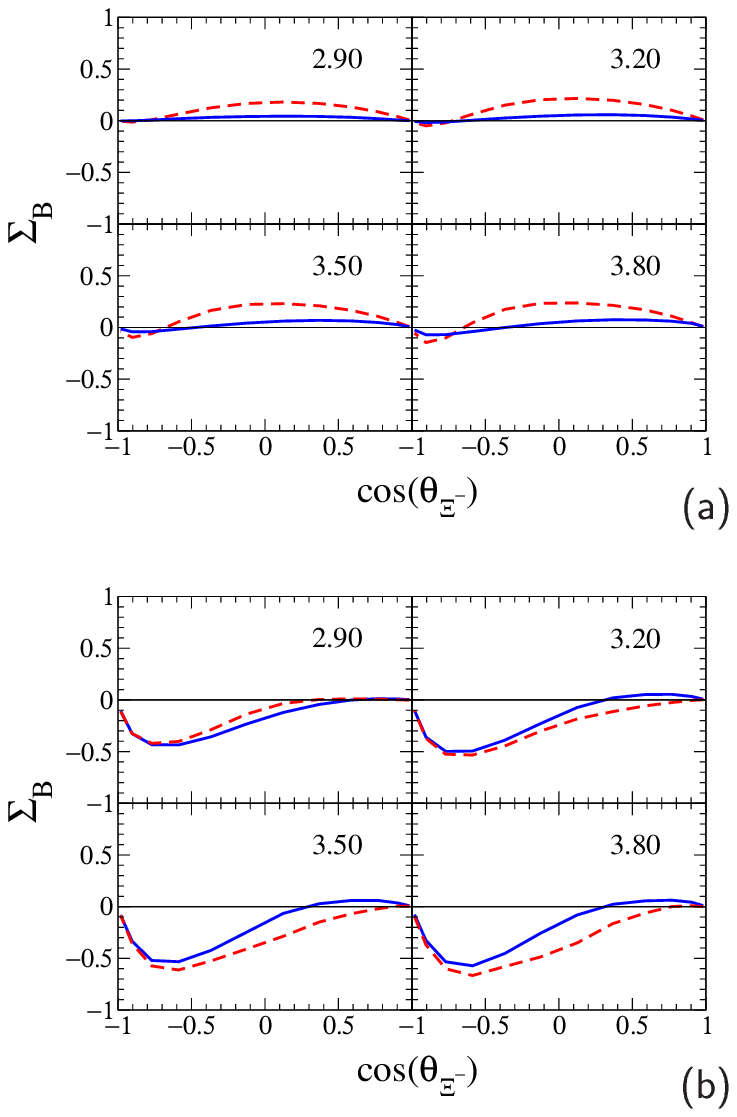}
\caption{\label{fig:SigmaB}%
(Color online)
Photon asymmetry as a function of the $\Xi^-$ emission angle in the
\mbox{c.m.} frame in $\gamma p \to K^+ K^+ \Xi^-$ for the models discussed
in (a) Sec.~\ref{sec:lowmass} (with only low-mass hyperons) and (b)
Sec.~\ref{sec:highmass} [with, in addition, the higher-mass $\Lambda(1800)$
and $\Lambda(1890)$ hyperons], respectively.
The solid and dashed curves correspond to the predictions of the model with
the pv- and ps-coupling choices, respectively. }
\end{figure}

Our predictions for the photon beam asymmetry,
\begin{equation}
\Sigma_B \equiv \frac{\sigma(\lambda=+1) - \sigma(\lambda=-1)}
                     {\sigma(\lambda=+1) + \sigma(\lambda=-1)} \ ,
\label{eq:Sigma_B}
\end{equation}
where $\sigma(\lambda)$ denotes the cross section with the linear photon
polarization along the $y$-axis ($\lambda=+1$) and $x$-axis ($\lambda=-1$), are
shown in Fig.~\ref{fig:SigmaB} as a function of the $\Xi^-$ emission angle. The
solid and dashed curves correspond to the predictions of the model in the
present work with the pv- and ps-coupling choices, respectively.
In Fig.~\ref{fig:SigmaB}(b), we display the results
of the model of the present subsection.
First, we see that both curves are practically the same.
Second, they are largely negative at backward angles, a feature that becomes
more pronounced as the photon energy increases.
This is a characteristic feature of the $t$-channel $K$-exchange mechanism.
As noted before, $t$-channel processes contribute mostly at small $t$ and
higher incident energies.
In the \mbox{c.m.} frame, this implies a backward-angle emission of the
$\Xi^-$ hyperon. Now, the beam asymmetry due to the $t$-channel
$K$-exchange corresponding to Fig.~\ref{fig:diagram}(a) alone is identical to
$\Sigma_B = -1$, since the three-momentum ${\bf q}_1^{}$ of the emitted $K^+$
can be chosen to be in the $xz$-plane without loss of generality.
Given in Fig.~\ref{fig:SigmaB}(a) are the corresponding predictions of the
model discussed in the previous subsection, Sec.~\ref{sec:lowmass}, where
the radiative transition diagram [Fig.~\ref{fig:diagram}(d) with $Y\ne Y'$]
dominates.
We see that they are small and positive for most $\Xi^-$ emission angles,
which is very different from the model results of the present subsection
[Fig.~\ref{fig:SigmaB}(b)].
This, therefore, demonstrates that the beam asymmetry can be used as another
independent observable to identify the relevant production mechanisms, i.e.,
whether the $t$-channel $K$ exchanges dominate or the radiative transition
diagrams dominate.

%

\begin{figure}[b!]\centering
\includegraphics[width=0.845\columnwidth,angle=0,clip]{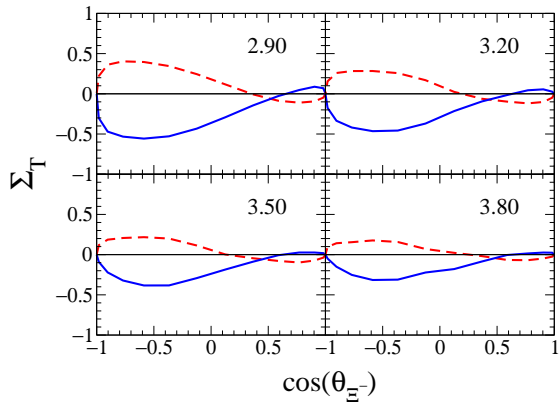}
\hspace{2mm}\mbox{}
\caption{\label{fig:SigmaT}%
(Color online)
Target asymmetry as a function of the $\Xi^-$ emission angle in
the \mbox{c.m.} frame in $\gamma p \to K^+ K^+ \Xi^-$.
The solid and dashed curves correspond to the predictions of the model in
Sec.~\ref{sec:highmass} for the pv- and ps-coupling choices, respectively.}
\end{figure}

Here, it should be noted that, in general, spin observables are much more
sensitive to the details of theoretical models, in particular, to the
background FSI effects.
As mentioned in Sec.~I, the latter is not considered in the present work.
However, to the extent that the kaons couple strongly to the (high-mass)
spin-1/2 hyperon resonances, the features just discussed above for the beam
asymmetry should hold.%
\footnote{Note that the pole part of the FSI is accounted for by the
resonances.}
The same observation applies for the target asymmetry discussed below.
For all the observables considered so far in the present work, the pv- and
ps-coupling schemes give very similar results to each other, although the
dynamical contents can differ substantially from each other.
In particular, the dominant contribution to the total cross sections can
arise either from the spin-1/2 or spin-3/2 hyperons depending on the choice
of the pv- or ps-coupling at the $BYK$ vertex for spin-1/2 baryon $B$ and
hyperon $Y$ (cf.\ Figs.~\ref{fig:txsc_lmR} and \ref{fig:txsc_hmR}).
In an effort to distinguish between the two coupling schemes, we have
computed the target asymmetry which is defined as
\begin{equation}
\Sigma_T \equiv \frac{\sigma(\lambda_N^{}=+1) - \sigma(\lambda_N^{}=-1)}
                     {\sigma(\lambda_N^{}=+1) + \sigma(\lambda_N^{}=-1)} \ ,
\label{eq:Sigma_T}
\end{equation}
and found that it can provide a tool for testing these coupling schemes.
Here, $\sigma(\lambda_N=\pm 1)$ denotes the cross section with the
polarized target nucleon along the $\pm y$-axis.
In Fig.~\ref{fig:SigmaT}, the predictions of this subsection's model for the
target asymmetry are shown.
As can be seen, the predictions corresponding to the pv-coupling
(solid curves) differ markedly from those corresponding to the ps-coupling
(dashed curves) at backward angles.
Note that the sensitivity to the ps-pv mixing parameter at backward angles is
due to the strong $t$-channel $K$-exchange contribution and arises from the
fact that the ps-coupling involves $\gamma_5$ while the pv-coupling
involves $\gamma_5 q_\mu \gamma^\mu$ at the meson-baryon vertex.
The latter leads to an amplitude which contains an extra
$\bm{\sigma}\cdot{\bf q}$ factor as compared to the former choice.
Therefore, the target asymmetry offers a potential means of distinguishing
the two types of couplings at the $BYK$ vertex.
The larger difference between the pv- and ps-coupling at low energies arises
from the $\Lambda(1800)1/2^-$ resonance contribution which is just below
the threshold energy.

%

\subsection{Higher-mass resonances (II)}

In the remaining part of this Section we further explore the influence of the
undetermined parameters in the present work.
Specifically, in this subsection we consider the negative signs for the
parameters mentioned in item b) of Sec.~II.
In this discussion, we restrict ourselves to the pv-coupling choice.

\begin{figure}[t!]\centering
\includegraphics[width=0.85\columnwidth,angle=0,clip]{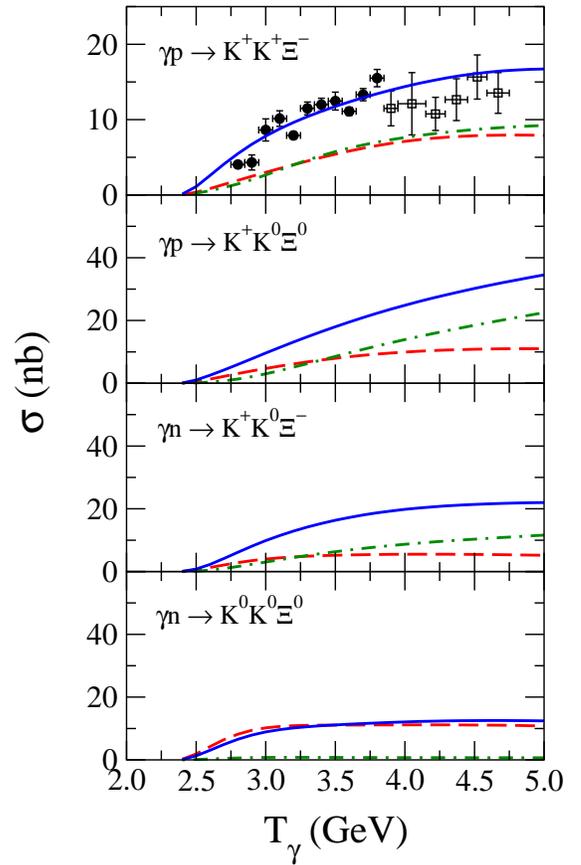}
\caption{\label{fig:txsc_hmR_2}%
(Color online)
Same as Fig.~\ref{fig:txsc_hmR}(a), but with the signs of the
coupling constants mentioned in item b) of Sec.~II chosen to be negative.}
\end{figure}

\begin{figure}[t!]\centering
\includegraphics[width=0.85\columnwidth,angle=0,clip]{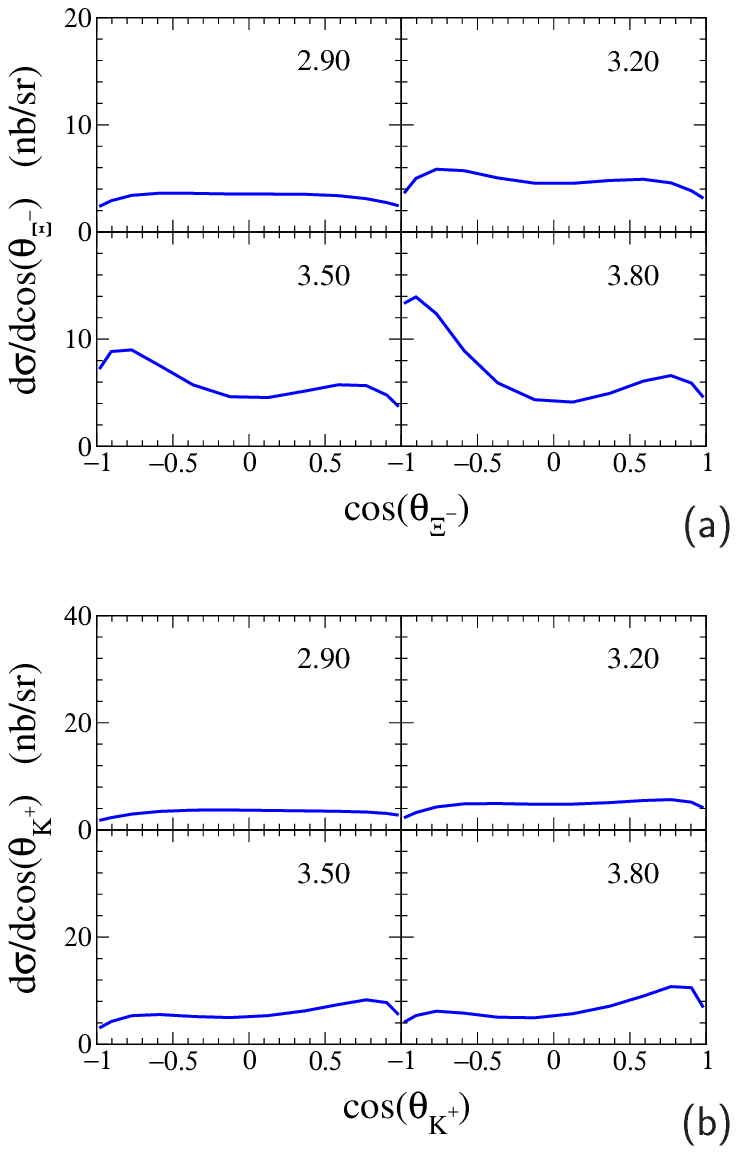}
\caption{\label{fig:dxsc_hmR_2}%
(Color online)
Same as Fig.~\ref{fig:dxsc_hmR}, but for the results of
Fig.~\ref{fig:txsc_hmR_2}.}
\end{figure}

Presented in Fig.~\ref{fig:txsc_hmR_2} are the results for the total cross
sections.
The cutoff parameter $\Lambda_B=1250$ MeV and the exponent $n \to \infty$
have been readjusted to reproduce the preliminary total cross section data
in the $\gamma p \to K^+ K^+ \Xi^-$ reaction channel.
The baryonic form factor in Eq.~(\ref{eq:ffB}) then corresponds to a
Gaussian form with the width of $1250$ MeV.
In addition, we have also adjusted slightly the mass of the two
higher-mass hyperons to be $m_{\Lambda 1/2^-}=1850$ and
$m_{\Lambda 3/2^+}=1950$ MeV to reproduce the total cross section data.
Note that the PDG masses for the $\Lambda(1800)$ and $\Lambda(1890)$ are
in the range of $1720 \sim 1850$ MeV and $1850 \sim 1910$ MeV, respectively.
The product of the coupling constants
$g_{N\Lambda K}^{}g_{\Xi\Lambda K}^{} = 2$ for both the
$\Lambda(1850)1/2^-$ and $\Lambda(1950)3/2^+$ hyperons have been kept as in
the previous subsection so as to result in the similar $\Xi^-$ and $K^+$
angular distributions as in Fig.~\ref{fig:dxsc_hmR}.
The corresponding angular distributions are displayed in
Fig.~\ref{fig:dxsc_hmR_2}.

%

%

\begin{figure}[t!]\centering
\includegraphics[width=0.866\columnwidth,angle=0,clip]{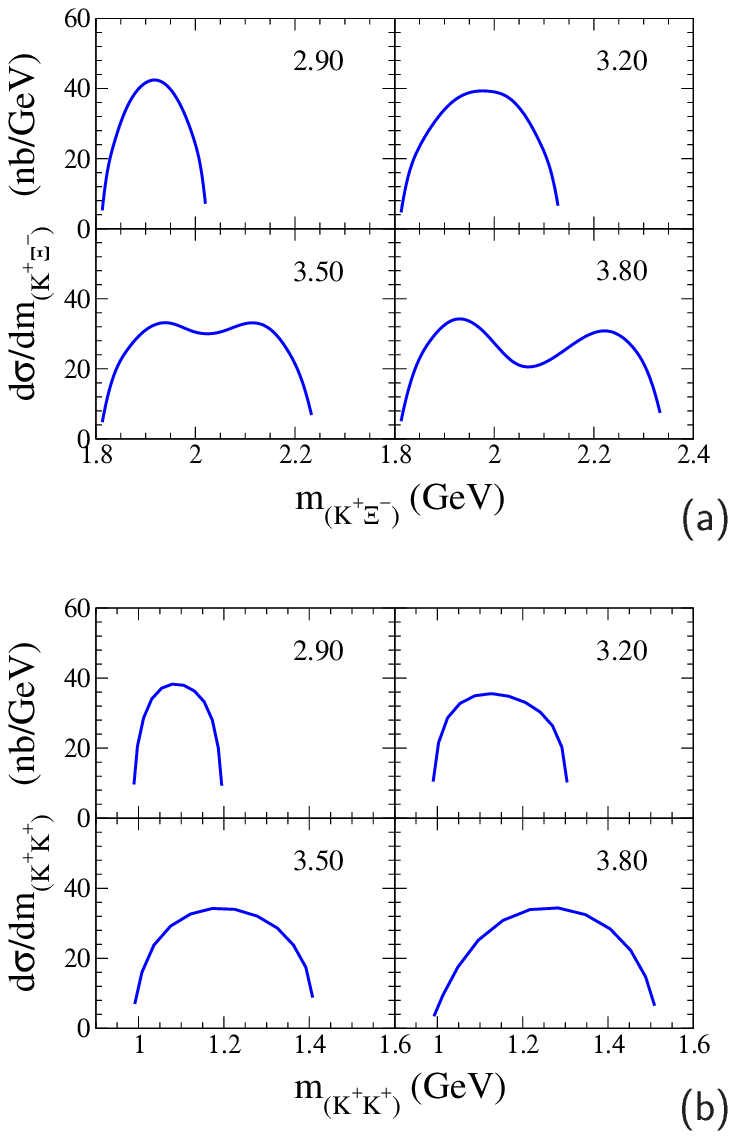}
\caption{\label{fig:dxsc_invmass_hmR_2}%
(Color online)
Same as Fig.~\ref{fig:dxsc_invmass_lmR}, but for the results of
Fig.~\ref{fig:txsc_hmR_2}.}
\end{figure}
%

\begin{figure}[b!]\centering
\includegraphics[width=0.41\textwidth,angle=0,clip]{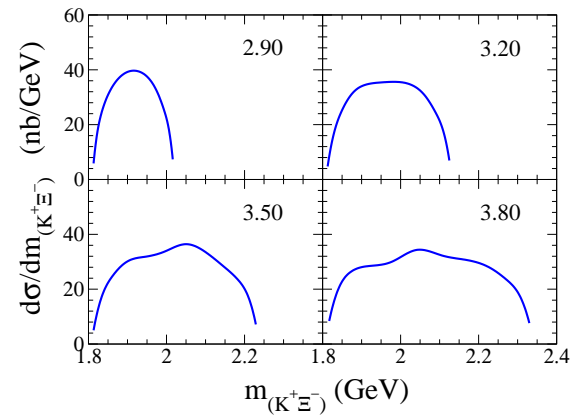}
\caption{\label{fig:dxsc_invmass_hmR_3R}%
(Color online)
Same as Fig.~\ref{fig:dxsc_invmass_hmR_2}, but with the additional
(fictitious) $\Lambda(2050)3/2^+$ resonance (see text for detailed
explanation).}
\end{figure}
%

Figure~\ref{fig:dxsc_invmass_hmR_2} presents the predictions for the
$K^+\Xi^-$ and $K^+K^+$ invariant mass distributions, respectively, in
$\gamma p \to K^+ K^+ \Xi^-$.
They correspond to the total cross section results in
Fig.~\ref{fig:txsc_hmR_2} and exhibit similar features to those
shown in Fig.~\ref{fig:dxsc_invmass_hmR} in the previous subsection,
although here the two bumps are more symmetric.

\subsection{Higher-mass resonances above 2 GeV}

The structure of the $K^+\Xi^-$ invariant mass distribution predicted in
Figs.~\ref{fig:dxsc_invmass_hmR} and \ref{fig:dxsc_invmass_hmR_2} may change
qualitatively if there is a significant contribution from some additional
resonances in the $2.0$ GeV mass region.
In fact, there are well established spin-5/2 and -7/2 $\Lambda$ and $\Sigma$
hyperons with masses around 2.05 GeV (cf.~Table~\ref{tbl:hyperons}) which may
potentially affect the $K^+\Xi^-$ invariant mass distribution in this energy
region.
It is conceivable, therefore, that these resonances may be capable to fill
up the valley between the two bumps in the predicted invariant mass
distribution seen on the left-hand sides of Figs.~\ref{fig:dxsc_invmass_hmR}
and \ref{fig:dxsc_invmass_hmR_2}.
However, as explained in the Introduction, since there is no detailed
dynamical information available about these resonances, their inclusion into
our model would be very speculative and an investigation of the detailed
effects of such higher-spin resonances should be left for future work.
However, the question as to whether some resonance in the 2-GeV region can
fill the valley can be addressed qualitatively by simulating high-spin
resonances by a \emph{fictitious} hyperon resonance in this energy region.
Concretely, therefore, in addition to the known $\Lambda(1800)1/2^-$ and
$\Lambda(1890)3/2^+$ states (and  the low-mass hyperons considered in the
previous subsections), we employ a (fictitious) $\Lambda(2050)3/2^+$ resonance
with $\Gamma_{\Lambda(2050)}=200$ MeV.
The product of the coupling
constants $g_{N\Lambda K}g_{\Xi\Lambda K} = 2$ for $\Lambda(1800)$,
$g_{N\Lambda K}^{}g_{\Xi\Lambda K}^{} = 1.2$ for $\Lambda(1890)$, and
$g_{N\Lambda K}^{}g_{\Xi\Lambda K}^{} = (1.2)^2$ for $\Lambda(2050)$
are used, in conjunction with the cutoff parameter $\Lambda_B=1250$ MeV and
$n\to\infty$ in order to reproduce the measured total cross
section~\cite{Guo} and keep the shape of the $\Xi^-$ and $K^+$ angular
distributions backward- and forward-peaked, respectively, as in
Fig.~\ref{fig:dxsc_hmR_2}.
We refrain from showing the corresponding total cross section and angular
distribution results here because they are practically the same as those
shown in Figs.~\ref{fig:txsc_hmR_2} and \ref{fig:dxsc_hmR_2}.

As can be seen in Fig.~\ref{fig:dxsc_invmass_hmR_3R}, the bump structures
indeed have disappeared completely due to the $\Lambda(2050)$ resonance
whose contribution fills up the valley in the $K^+\Xi^-$ invariant mass
around $m_{K^+\Xi^-}=2$ GeV seen in Figs.~\ref{fig:dxsc_invmass_hmR} and
\ref{fig:dxsc_invmass_hmR_2}.
We may expect, therefore, that a similar effect will also occur when the known
higher-spin resonances in this energy range are considered.

These considerations show that the study of $\Xi$ photoproduction may
provide useful information about higher-mass hyperon resonances which makes
it all the more desirable to have more rigorous investigations that provide
a better understanding of the dynamics of such higher-mass and higher-spin
resonances.

\section{Summary}

In summary, we have explored the reaction $\gamma N \to K K \Xi$ within a
relativistic meson-exchange model of hadronic interactions.
This is the first theoretical investigation of this reaction in connection
with the cascade spectroscopy program initiated recently by the
CLAS Collaboration at JLab~\cite{Price1,Price2,Price3,Guo}.

Most of the parameters of the model involving the low-mass hyperons in the
intermediate states were determined from the empirical data and/or quark models
and SU(3) symmetry considerations.
It is found that the dominant reaction mechanism arising from those low-mass
hyperons is the radiative transitions
[cf.\ Fig.~\ref{fig:diagram}(d) with $Y\ne Y'$], especially, for producing the
$\Xi^-$ in the $\gamma p \to K^+ K^+ \Xi^-$ reaction, which is currently under
investigation by the CLAS Collaboration at JLab~\cite{Guo}.
This production mechanism is also found to lead to a forward- (backward-)
peaked angular distribution of the produced $\Xi^-$ ($K^+$).

We have also explored the possible influence of the higher-mass hyperons in
$\gamma N \to K K \Xi$, which are expected to contribute more significantly to
this reaction than the low-mass hyperons because they are energetically more
favored than the latter resonances. The difficulty in quantifying the
contributions from these high-mass hyperons is the complete lack of information
about the strengths of their couplings to the cascade baryons. Nevertheless, we
have shown that these high-mass resonances may lead to a dominance of the
$t$-channel $K$-exchange mechanism for producing the cascade baryons. Moreover,
the angular distribution of the produced $\Xi^-$ and the photon asymmetry in
$\gamma p \to K^+ K^+ \Xi^-$ were shown to offer two independent ways of
possibly distinguishing between the $t$-channel $K$-exchange and the radiative
transitions as the dominant mechanisms for $\Xi^-$ photoproduction. The target
asymmetry was also shown to possibly impose constraints on the ps-pv mixing
parameter at the $BYK$ vertex involving spin-1/2 baryons.

In addition, the $K^+\Xi^-$ invariant-mass distribution in $\gamma p \to K^+
K^+ \Xi^-$ yields a clear information on the high-mass hyperon resonance
contributions. However, care must be taken to avoid misidentifying the second
bump structure, which is kinematic in origin, with the formation of a
higher-mass resonance during the reaction. Moreover, the non-existence of the
bump structure in the $K^+\Xi^-$ invariant mass distribution may happen due to
the overlap of broad resonances as we have explicitly shown in
Fig.~\ref{fig:dxsc_invmass_hmR_3R}.
These findings show that $\Xi$ photoproduction is indeed well suited for
investigating the properties of higher-mass hyperon resonances.

We conclude from the present work that one needs to consider concomitantly
not only the total cross sections and their angular distributions but also
other observables, such as invariant-mass distributions, beam asymmetries,
and target asymmetries, in order to learn about the cascade photoproduction
reaction.
This is required, especially when the reaction mechanism is unclear as in the
current case. These observables have different role in identifying the
production mechanisms.

Finally, the present effort is just a first step toward building a more
complete and realistic model for describing cascade baryon photoproduction off
nucleons.
Our findings should be useful for future investigation of this reaction
both experimentally and theoretically.
Experimentally, we are aware of that the CLAS Collaboration is currently
analyzing the $\gamma p \to K^+ K^+ \Xi^-$ reaction and extracting the
angular distributions of $\Xi^-$ and $K^+$ as well as the $K^+\Xi^-$ and
$K^+K^+$ invariant-mass distributions~\cite{Guo1} in addition to the total
cross sections.
As we have shown, these observables will certainly be very important in
learning about this reaction.
Also, it would be very interesting to have the beam and target asymmetries
measured in future experiments.
Studying the reaction channels other than $\gamma p \to K^+ K^+ \Xi^-$ is
also required, for these will help disentangle the isoscalar $\Lambda$ and
isovector $\Sigma$ hyperon contributions.
Theoretically, we should investigate the effects of the other
higher-mass baryon resonances, not only those of the
rather well-established $J^P=1/2^-$ and -$3/2^+$ resonances that have been
neglected in the present study but, especially, those of higher-spin
(spin-5/2 and -7/2) resonances (cf. Table~\ref{tbl:hyperons}).
Furthermore, in future works, one should investigate the $K\Xi$ final-state
interaction and other effects mentioned in the Introduction which were not
considered in the present study due to the present lack of detailed
information about the relevant reaction dynamics.
Evidently, the production of $\Xi(1530)$ should be considered as a next
step in the cascade spectroscopy program.

\acknowledgments

We are indebted to L.~Guo, D.~P. Weygand, and the CLAS Collaboration for
providing us with the preliminary data on $\gamma p \to K^+ K^+ \Xi^-$.
We also thank L.~Guo, D.~P. Weygand, J.~Price, and T.-S.~H. Lee for many
fruitful discussions.
This work was supported in part by the COSY Grant \mbox{No.} 41445282 (COSY-58).

\appendix*


\section{}

The interaction Lagrangians used to construct our model for the production
amplitudes shown in Fig.~\ref{fig:diagram} are given in this Appendix.
For further convenience, we define the operators
\begin{equation}
\hat\Gamma^{(+)} = \gamma_5 \qquad\text{and}\qquad \hat\Gamma^{(-)} = 1.
\end{equation}

The following Lagrangians describe the hadronic vertices:

\textit{$BYK$ Lagrangian} ($B=$ spin-1/2 baryon, $Y=$ spin-1/2 hyperon):
\begin{eqnarray}
{\cal L}^{(\pm)}_{BYK} &=& \mp g_{BYK}^{} \nonumber \\
&\times & \left[  i\lambda \bar{K}\, \bar Y \, \hat\Gamma^{(\pm)} +
\frac{1 - \lambda}{m_Y\pm m_B^{}}\, (\partial_\mu \bar{K}) \bar Y \,
\hat\Gamma^{(\pm)} \gamma^\mu \right] B
\nonumber \\
& + & \hc,
\label{BYK12}
\end{eqnarray}
where $B$, $Y$, and $K$ stand for the baryon, hyperon, and kaon fields,
respectively.
The upper and lower signs refer to whether $B$ and $Y$ have the same parity
($+$) or the opposite parity ($-$).
The masses $m_Y$ and $m_B$ are those of the hyperon $Y$ and baryon $B$,
respectively, and the ps-pv mixing parameter is denoted by $\lambda$.
For an isovector hyperon, $\bar Y\to\bar{\bm{Y}}\cdot\bm{\tau}$
$(Y\to\bm{\tau}\cdot\bm{Y})$ if $B$ is an isospin-1/2 baryon.
If $B$ is an isospin-3/2 baryon instead,
$\bar Y\to\bar{\bm{Y}}\cdot\bm{T}^\dagger$ $(Y\to\bm{T}\cdot\bm{Y})$,
where $\bm{T}$ is the isospin transition operator whose definition may be
found elsewhere, e.g., in Ref.~\cite{ONL04}.

\textit{$BYK^*$ Lagrangian} ($B=$ spin-1/2 baryon, $Y=$ spin-1/2 hyperon):
\begin{eqnarray}
{\cal L}^{(\pm)}_{BYK^*} &=& - g_{BYK^*}^{} \bar B \hat\Gamma^{(\mp)}
\nonumber \\
&& \mbox{} \times
\left( \gamma^\mu Y K^*_\mu - \frac {\kappa_{BYK^*}^{}}{m_N^{}}
\sigma^{\mu\nu} Y \partial_\nu K^*_\mu \right) +\hc .
\nonumber \\
\label{BYK*12}
\end{eqnarray}

\textit{$BYK$ Lagrangian} ($B=$ spin-1/2 baryon, $Y=$ spin-3/2 hyperon):
\begin{equation}
\mathcal{L}^{(\pm)}_{BYK} = \frac{f_{BYK}^{}}{m_K^{}}
\left(\partial^\nu \bar{K}\right)
{\bar Y_\nu} \hat\Gamma^{(\pm)} B +\hc,
 \label{BYK32}
\end{equation}
where $m_K$ stands for the kaon mass.

\textit{$BYK^*$ Lagrangian} ($B=$ spin-1/2 baryon, $Y=$ spin-3/2 hyperon):
\begin{eqnarray}
\mathcal{L}^{(\pm)}_{BYK^*} &=& i \frac{g^{(1)}_{BYK^*}}{2m_N^{}}
\bar{F}^{\mu\nu}
{\bar Y_\nu} \gamma_\mu \hat\Gamma^{(\pm)} B
\nonumber \\ && \mbox{}
   - \frac{g^{(2)}_{BYK^*}}{(2m_N^{})^2} \bar{F}^{\mu\nu}
(\partial_\mu {\bar Y_\nu}) \hat\Gamma^{(\pm)} B  + \hc,
\nonumber \\
 \label{BYK*32}
\end{eqnarray}
where $\bar{F}^{\mu\nu} = \partial^\mu \bar{K}^{*\nu} - \partial^\nu
\bar{K}^{*\mu}$ and $m_N^{}$ is the nucleon mass.

\textit{$BYK$ Lagrangian} ($B=$ spin-3/2 baryon, $Y=$ spin-3/2 hyperon):
\begin{equation}
\mathcal{L}^{(\pm)}_{BYK} = \frac{f_{BYK}^{}}{m_K^{}}
\bar{B}^\nu \gamma_\mu \gamma_5 Y_\nu
\partial^\mu K
+ \hc.
 \label{BYK33}
\end{equation}

The electromagnetic vertices are derived from the following Lagrangian
densities.

\textit{$BB\gamma$ Lagrangian} ($B=$ spin-1/2 baryon):
\begin{equation}
{\cal L}_{BB\gamma} = - \bar B \left[\left(e_B^{} \gamma^\mu -
e\kappa_B^{}\frac{\sigma^{\mu\nu}\partial_\nu}{2m_N^{}}
\right) A_\mu \right] B ,
\label{BBgamma}
\end{equation}
where $A_\mu$ stands for the photon field and $e_B^{}$ is the charge operator
of the baryon while $e$ stands for the elementary charge unit.
The baryon's anomalous magnetic moment is given by $\kappa_B^{}$ in units
of nuclear magneton.

\textit{$YY'\gamma$ transition Lagrangian} ($Y=$ spin-1/2 hyperon, $Y'=$
spin-1/2 hyperon):
\begin{equation}
{\cal L}^{(\pm)}_{YY'\gamma}  =  e\frac {\kappa_{YY'\gamma}^{}}{2m_N^{}}\,
\bar Y' \hat\Gamma^{(\mp)} \sigma_{\mu\nu}(\partial^\nu A^\mu) Y +\hc.
\label{BYgamma}
\end{equation}

\textit{$YY'\gamma$ Lagrangian} ($Y=$ spin-1/2 hyperon, $Y'=$ spin-3/2
hyperon):
\begin{eqnarray}
\mathcal{L}^{(\pm)}_{BY\gamma} &=& i \frac{e g_1^{}}{2m_N^{}} A^{\mu\nu}
{\bar Y'_\nu} \gamma_\mu \hat\Gamma^{(\pm)}  Y
\nonumber \\ && \mbox{}
- \frac{e g_2^{}}{(2m_N^{})^2} A^{\mu\nu}
(\partial_\mu {\bar Y'_\nu})  \hat\Gamma^{(\pm)} Y  + \hc,
 \label{BYgamma:32}
\nonumber \\
\end{eqnarray}
where $A^{\mu\nu} = \partial^\mu A^\nu - \partial^\nu A^\mu$.

\textit{$BB\gamma$ Lagrangian} ($B=$ spin-3/2 baryon):
\begin{eqnarray}
{\cal L}_{BB\gamma} & = & \bar B^\mu e_B^{} \gamma_\alpha
\left\{g_{\mu\nu}
- \frac{1}{2}(\gamma_\mu\gamma_\nu + \gamma_\nu\gamma_\mu)\right\}
A^\alpha  B^\nu
\nonumber \\ && \mbox{}
 - e\bar B^\mu \kappa_B^{}
\frac{\sigma^{\alpha\nu}(\partial_\nu A_\alpha)}{2m_N^{}}
B_\mu .
\label{BBgamma:33}
\end{eqnarray}

\textit{$BYK\gamma$ Lagrangian} ($B=$ spin-1/2 baryon; $Y=$ spin-1/2 hyperon):
\begin{eqnarray}
{\cal L}_{BYK\gamma} &=&  -e g_{BYK}^{}\frac{1-\lambda}{m_Y^{}\pm m_B^{}}
K (\bm{I} \times \bar {\bm{Y}})_3 \hat\Gamma^{(\pm)}\gamma_\mu A^\mu B
\nonumber \\ && \mbox{}
+ \hc,
\label{BYKgamma}
\end{eqnarray}
where $\bm{I} = \bm{\tau}$ or $\bm{T}$ if $B$ is an isospin-1/2 or -3/2 baryon.

\textit{$BYK\gamma$ Lagrangian} ($B=$ spin-1/2 baryon; $Y=$ spin-3/2 hyperon):
\begin{equation}
{\cal L}_{BYK\gamma} = -e \frac{f_{BYK}^{}}{m_K^{}} K
(\bar {\bm{Y}}^\mu \times \bm{I}^\dagger)_3
 A^\mu B + \hc.
\label{BYKgamma:32}
\end{equation}

In addition, the $KK\gamma$ and $KK^*\gamma$ interactions are
\begin{equation}
{\cal L}_{KK\gamma} = ie \left( K^-(\partial_\mu K^+) - (\partial_\mu K^-) K^+\right) A^\mu,
\label{KKgamma}
\end{equation}
and
\begin{align}
{\cal L}_{KK^*\gamma} & =
 e\frac{g^0_{KK^*\gamma}}{m_K^{}}
\varepsilon^{\mu\nu\alpha\beta}\partial_\mu A_\nu
 \nonumber \\
&\quad\mbox{} \times
\Big[ (\partial_\alpha K^{*0}_\beta) \bar K^0 +
(\partial_\alpha \bar K^{*0}_\beta) K^0\Big] \nonumber \\
&\quad\mbox{}
  +  e\frac{g^c_{KK^*\gamma}}{m_K^{}}
\varepsilon^{\mu\nu\alpha\beta}\partial_\mu A_\nu \nonumber \\
&\quad\mbox{}
 \times  \Big[ (\partial_\alpha K^{*-}_\beta) K^+ +
(\partial_\alpha K^{*+}_\beta) K^-\Big],
 \label{KK*gamma}
\end{align}
respectively.

The coupling constants in the above interaction Lagrangians are given in
Tables~\ref{tbl:cc} and \ref{tbl:trcc}. We use the propagators for spin-1/2 and
3/2 resonances introduced in Ref.~\cite{NH}, which are consistent with the
above interaction Lagrangians and the Ward-Takahashi identity.
Specifically, the propagator for a spin-1/2
resonance, with mass $m_R^{}$ and width $\Gamma$, reads
\begin{equation}
S_{1/2}(p) = \frac{1}{p\!\!\!/ - m_R^{} + \frac{i}{2}\Gamma},
\end{equation}
and that for a spin-3/2 resonance is
\begin{equation}
S_{3/2}(p) = \left[ (p\!\!\!/ - m_R^{}) g - i \frac{\Delta}{2} \Gamma
\right]^{-1} \Delta,
\end{equation}
where all indices are suppressed.
Here, $g\equiv g^{\mu\nu}$ is the metric tensor and
\begin{equation}
\Delta \equiv \Delta^{\mu\nu} =
-g^{\mu\nu} + \frac13 \gamma^{\mu} \gamma^\nu + \frac23
\frac{p^\mu p^\nu}{m_R^2} + \frac{\left( \gamma^\mu p^\nu - p^\mu \gamma^\nu
\right)}{3m_R^{}}.
\end{equation}
The pseudoscalar and vector meson propagators are
\begin{eqnarray}
\Delta_0(q) & = & \left( q^2 - m_p^2 \right)^{-1}  , \nonumber \\
D^{\mu\nu}(q) & = & \left( \frac{-g^{\mu\nu}
+ q^\mu q^\nu/m_v^2}{q^2 - m_v^2} \right) ,
\end{eqnarray}
respectively, where $m_p^{}$ denotes the mass of the pseudoscalar meson and
$m_v^{}$ the vector meson mass.

Our model contains form factors at the hadronic and electromagnetic
vertices to account for the composite nature of the hadrons.
Currently, no theory is available to calculate these form factors from more
basic principles.
Here, they are introduced phenomenologically.
We use the separable form for the form factor at the
baryon-baryon-meson vertex,
\begin{equation}
F(p'^2, p^2, q^2) = f_{B}^{}(p'^2) f_B^{}(p^2) f_M^{}(q^2)~,
 \label{eq:ffB'BM}
\end{equation}
where $p'^2$ and $p^2$ denote the momentum of the outgoing and incoming
baryon, respectively, and, $q^2$ is the momentum of the meson at the
hadronic vertex.
The function $f_B^{}$ is given by
\begin{equation}
f_B^{}(p^2) = \left( \frac{n\Lambda_B^4}{n\Lambda_B^4 + (p^2 - m_B^2)^2}
\right)^n  ,
\label{eq:ffB}
\end{equation}
where $m_B^{}$ denotes the mass of the baryon.
The cutoff parameter $\Lambda_B$ and $n$ are kept the same for all the
baryons in order to minimize the number of parameters and they are
treated as free parameters to be adjusted.
The above form of the form factor reduces to a Gaussian function with
a width $\Lambda_B$ in the limit of $n\to \infty$.
The function $f_M^{}$ is given by
\begin{equation}\label{eq:ffM}
\begin{aligned}
f_K^{}(q^2) &= \frac{\Lambda_K^2 - m_K^2}{\Lambda_K^2 -  q^2}  ,
 \\
f_{K^*}^{}(q^2) &= \exp\left( \frac{q^2 - m_{K^*}^2}{\Lambda_{K^*}^2}
\right),
\end{aligned}
\end{equation}
for pseudoscalar ($M=K$) and vector ($M=K^*$) meson, respectively.
Here, the cutoff parameters are fixed to be $\Lambda_K=1300$ MeV and
$\Lambda_{K^*}=1000$ MeV.%
\footnote{We have explored the sensitivity of the results to the cutoff
parameters $\Lambda_M$ ($M=K,K^*$) and found that they are not as
sensitive as to the parameters ($\Lambda_B,n$) of the baryonic
form factor.}
The Gaussian form factor for the vector meson $K^*$ is
motivated by the work of Ref.~\cite{Nijmegen1}, where the same form has
been employed in a three-dimensional approach.
However, other forms such as the dipole form factor can be employed as well
without changing the results qualitatively.

Keeping gauge invariance of the reaction amplitude is not a trivial task
when phenomenological hadronic form factors are present.
Here, gauge invariance is maintained through phenomenological contact
currents [diagrams (f) and (g) in Fig.~\ref{fig:diagram}] based on the
Ward-Takahashi identity by extending the method of Ref.~\cite{HH} for
one-meson photoproduction to two-meson photoproduction processes.
Explicitly, they are given by
\begin{equation}
C^\mu_1 = \Gamma^\mu_{c\ 1} (e_i^{}\tilde R_{s_1} - e_B^{}\tilde R_1)
+ \Gamma_1 \tilde C^\mu_1  ,
\label{eq:cnt1}
\end{equation}
where
\begin{eqnarray}
\tilde C^\mu_1 &=& -e_1^{}\frac{(2q_1-k)^\mu}{t_1-q_1^2}
\left(\tilde R_{t_1}-\hat F_1\right)
\nonumber \\ && \mbox{}
-e_i^{}\frac{(2p+k)^\mu}{s_1-p^2}\left(\tilde R_{s_1}-\hat F_1\right)
\nonumber \\ && \mbox{}
-e_B^{}\frac{(2p-2q_1+k)^\mu}{u_1-s_2}\left(\tilde R_1 -\hat F_1\right) ,
\label{eq:cnt1a}
\end{eqnarray}
with
\begin{equation}
\hat F_1 = \hat R_1 + \frac{1}{\hat R_1^2}
(\delta_1\tilde R_{t_1}-\hat R_1) (\delta_i\tilde R_{s_1}-\hat R_1)
(\delta_B\tilde R_1    -\hat R_1)
\label{eq:cnt1aa}
\end{equation}
(which corresponds to an off-shell generalization of Ref.~\cite{DW}) and
\begin{equation}
C^\mu_2 = \Gamma^\mu_{c\ 2} (e_B^{}\tilde R_2 - e_f^{}\tilde R_{u_2})
+ \Gamma_2 \tilde C^\mu_2 ,
\label{eq:cnt2}
\end{equation}
where
\begin{eqnarray}
\tilde C^\mu_2 &=& -e_2^{}\frac{(2q_2-k)^\mu}{t_2-q_2^2}
\left(\tilde R_{t_2}-\hat F_2\right)
\nonumber \\   && \mbox{}
-e_f^{}\frac{(2p'+k)^\mu}{u_2-p'^2}\left(\tilde R_{u_2}-\hat F_2\right)
\nonumber \\   && \mbox{}
-e_B^{}\frac{(2p'-2q_2+k)^\mu}{s_2-u_1}\left(\tilde R_2 -\hat F_2\right) ,
\label{eq:cnt2a}
\end{eqnarray}
with
\begin{equation}
\hat F_2 = \hat R_2 + \frac{1}{\hat R_2^2}(\delta_2\tilde R_{t_2}-\hat R_2)
(\delta_f\tilde R_{u_2}-\hat R_2) (\delta_B\tilde R_2 - \hat R_2)  .
\label{eq:cnt2aa}
\end{equation}
In Eq.~(\ref{eq:cnt1}), $\Gamma^\mu_{c\ 1}$ and $\Gamma_1$ denote the bare
$NYK\gamma$ and $NYK$ vertex, respectively.
Likewise, in Eq.~(\ref{eq:cnt2}), $\Gamma^\mu_{c\ 2}$ and $\Gamma_2$ denote
the bare $\Xi YK\gamma$ and $\Xi YK$ vertex, respectively.
They are obtained from the interaction Lagrangians Eqs.~(\ref{BYK12}) and
(\ref{BYKgamma}) for spin-1/2 hyperons and Eqs.~(\ref{BYK32}) and
(\ref{BYKgamma:32}) for spin-3/2 hyperons.
Here, $e_i^{}$, $e_f^{}$, $e_1^{}$, and $e_2^{}$ denote the combined
charge-isospin operators of the nucleon in the initial state, cascade, and
the kaons 1 and 2 in the final state, respectively, and
$e_B^{}=e_f^{}+e_2^{}=e_i^{}-e_1^{}$.
(Up to an isospin-dependent factor, the $e_x^{}$ are effectively the charges
of the respective particles; see Ref.~\cite{HH} for details.)
For non-zero charges $e_x^{}$, one has $\delta_x = 1$, and zero otherwise.
$\hat R$'s and $\tilde R$'s in the above equations are hadronic form
factors given by Eq.~(\ref{eq:ffB'BM}) in different kinematics.
We have
\begin{equation}
\hat R_1  =  F(s_2, p^2, q_1^2), \qquad \hat R_2  =  F(p'^2, u_1, q_2^2),
\label{ffRhat}
\end{equation}
and
\begin{align}
 \tilde R_1 &= F(u_1, p^2, q_1^2)~,
  & \tilde R_2  &=  F(p'^2, u_1, q_2^2)~,
  \nonumber\\
 \tilde R_{s_1} &= F(s_2, s_1, q_1^2)~,
  & \tilde R_{u_2}  &= F(u_2,u_1, q_2^2)~,
  \label{ffRtilde}
 \\ \tilde R_{t_1} &= F(s_2, p^2, t_1)~,
 & \tilde R_{t_2}  &=  F(p'^2, u_1, t_2)~,
 \nonumber
\end{align}
with
\begin{align}
s_1 &= (p+k)^2,   &  s_2  &=  (p'+q_2)^2, \nonumber \\
u_1 &= (p-q_1)^2, &  u_2  &=  (p'-k)^2, \label{ffkinem}\\
t_1 &= (q_1-k)^2, &  t_2  &=  (q_2-k)^2  .\nonumber
\end{align}
Here, $k$ and $p$ stand for the momenta of the photon and nucleon in the
initial state, while $q_1$, $q_2$ and $p'$, for the momenta of the two kaons
and cascade in the final state, respectively (see Fig.~\ref{fig:diagram}).
A detailed derivation of these contact currents will be reported
elsewhere~\cite{HN1}.

\end{document}